\documentclass[sigconf, balance=false]{acmart}

\renewcommand\footnotetextcopyrightpermission[1]{}%
\usepackage{acmart-taps}

\makeatletter
\def\@ACM@checkaffil{%
    \if@ACM@instpresent\else
    \ClassWarningNoLine{\@classname}{No institution present for an affiliation}%
    \fi
    \if@ACM@citypresent\else
    \ClassWarningNoLine{\@classname}{No city present for an affiliation}%
    \fi
    \if@ACM@countrypresent\else
        \ClassWarningNoLine{\@classname}{No country present for an affiliation}%
    \fi
}
\makeatother
\AtBeginDocument{%
  \providecommand\BibTeX{{%
    \normalfont B\kern-0.5em{\scshape i\kern-0.25em b}\kern-0.8em\TeX}}}

\setcopyright{acmcopyright}

\usepackage{float}
\usepackage[utf8]{inputenc}
\usepackage{url}
\usepackage{multirow}
\usepackage{array}
\usepackage{pifont}
\newcommand{\twoline}[1]{\begin{tabular}[c]{@{}c@{}}#1\end{tabular}}
\newcommand{\cmark}{\ding{51}}
\newcommand{\xmark}{\ding{55}}
\newcommand{\pmark}{\ding{51}\kern-0.45em\raisebox{0.45ex}{\scalebox{0.6}{\ding{55}}}}

\usepackage{amssymb}
\usepackage{natbib} 
\usepackage{graphicx}
\usepackage{color}
\usepackage{xcolor}
\usepackage{tikz}
\usetikzlibrary{positioning,fit,arrows.meta,backgrounds,calc,shapes.symbols,shapes.geometric}
\usepackage{verbatim}
\usepackage{algorithm, algorithmic}
\usepackage{amsmath}
\usepackage{xspace}

\newcommand{\ignore}[1]{{}}

\newcommand{\rev}[1]{#1}%

\usepackage{listings}
\definecolor{keyword-green}{rgb}{0.2, 0.5, 0.3}
\definecolor{string-red}{rgb}{0.7, 0.1, 0.2}
\definecolor{important-red}{rgb}{0.8, 0.25, 0.2}
\definecolor{comment-grey}{rgb}{0.4, 0.5, 0.6}
\definecolor{item-blue}{rgb}{0.25, 0.3, 0.85}
\definecolor{text-blue}{rgb}{0.25, 0.2, 0.4}
\definecolor{number-black}{rgb}{0.1, 0.1, 0.15}
\definecolor{another-orange}{rgb}{0.7, 0.5, 0.15}

\lstdefinestyle{sql-style}{
language=SQL,
commentstyle=\color{comment-grey},
keywordstyle=\color{keyword-green}\bfseries,
stringstyle=\color{string-red},
numbers=left,
numbersep=5pt,
numberstyle=\color{number-black}\tiny,
frame=lines,
breaklines=true,
basewidth={.5em},
morekeywords={FILTER},
moredelim=**[is][\color{item-blue}]{~}{~},
framextopmargin=0.5em,
framexbottommargin=0.5em,
}

\lstdefinestyle{py-style}{
language=Python,
commentstyle=\color{comment-grey},
keywordstyle=\color{keyword-green}\bfseries,
stringstyle=\color{string-red},
numbers=left,
numbersep=5pt,
numberstyle=\color{number-black}\tiny,
frame=lines,
breaklines=true,
basewidth={.5em},
morekeywords={FILTER},
moredelim=**[is][\color{item-blue}]{~}{~},
moredelim=**[is][\color{item-blue}\bfseries]{~~}{~~},
moredelim=[is][\color{string-red}]{--}{--},
framextopmargin=0.5em,
framexbottommargin=0.5em,
}

\lstdefinestyle{sparql-style}{
language=sparql,
commentstyle=\color{commend-grey},
keywordstyle=\color{keyword-green}\bfseries,
stringstyle=\color{string-red},
numbers=left,
numbersep=5pt,
numberstyle=\color{number-black}\tiny,
rulecolor=\color{number-black},
frame=lines,
breaklines=true,
basewidth={.5em},
morekeywords={FILTER},
moredelim=**[is][\color{item-blue}\bfseries]{~}{~},
moredelim=**[is][\color{keyword-green}\bfseries]{!}{!},
moredelim=**[is][\color{important-red}\bfseries]{=}{=},
moredelim=**[is][\color{important-red}]{==}{==},
moredelim=**[is][\bfseries]{!!}{!!},
framextopmargin=0.5em,
framexbottommargin=0.5em,
}

\usepackage{acmart-taps}

\begin{document}

\title{Context Is Not Free: Carbon-Aware LLM Routing for Multi-Turn LLM Web Services across Geo-Distributed Data Centers}

\author{Xiaoyang Zhang}
\affiliation{%
  \institution{Nanyang Technological University}
}

\author{Yuejun Yan}
\affiliation{%
  \institution{Alibaba Group}
}

\author{Qingang Zhang}
\affiliation{%
  \institution{Nanyang Technological University}
}

\author{Yonggang Wen}
\affiliation{%
  \institution{Nanyang Technological University}
}

\renewcommand{\shortauthors}{Zhang et al.}

\begin{abstract}
The sustainability of the Web has drawn growing attention, as large language model (LLM)-based web services leave a substantial carbon footprint. These services typically run in geo-distributed data centers whose grid carbon intensity varies across regions and over time, so routing each query to a cleaner region or a smaller model can reduce emissions. A conversation, however, is a stateful web session: each turn reuses the key--value (KV) cache of its history, which resides only where earlier turns were served. Serving a turn on another deployment requires recomputing this history, whose carbon cost grows with the conversation and can outweigh the saving of the move. Existing carbon-aware LLM routers treat a conversation's queries as independent and ignore this cost. We present CarbonRouter, an online framework that jointly selects the model, GPU type, and region for each query to minimize carbon emissions. CarbonRouter introduces a cache-aware carbon cost model, which charges history recomputation to every candidate deployment using cached-token counts from the serving backends, and uses a lightweight gate that keeps a session on its cached deployment when it would likely move back at the next turn and rebuild its history again. We evaluate CarbonRouter in trace-driven experiments on two real-world conversation datasets with five LLMs, three GPU types, and four regions, where it achieves the lowest carbon of all policies. CarbonRouter reduces CO$_2$ by 16.1\% on average relative to FairCarbon, a state-of-the-art carbon-aware router with the same action space, while also improving predicted answer quality and estimated latency. Moreover, the saving rises further to 23.1\% for sessions longer than 30 turns.
\end{abstract}

\keywords{Sustainability of Web, Carbon-aware Computing, LLM Routing}

\maketitle
\pagestyle{plain}%
\section{Introduction}
\label{sec:intro}

Large language model (LLM)-based web services, such as chat assistants, AI search engines, and coding assistants, now answer users through web interfaces and APIs, and every turn of a conversation is a web request that runs inference in a data center~\cite{jegham2025hungry}. Meanwhile, as platforms increasingly rely on LLMs as the front end of user interaction, the carbon footprint of model inference has drawn growing attention~\cite{jegham2025hungry,iea2025energyai}. Like other large Web services, an LLM-based web service typically runs in several geo-distributed data centers and routes each request among them \cite{liu2011greening}. This routing can cut operational carbon in two ways: serve a query with less energy (on a smaller model that is good enough or on a more efficient GPU), or in a region whose grid is currently cleaner, since grid carbon intensity differs across regions and changes within minutes.

Moving a query is not free. Most traffic of LLM-based web services consists of multi-turn conversations, and a conversation is a stateful web session. A turn is served with the key--value (KV) cache of the conversation history, which resides only on deployments that served earlier turns, until the serving backend evicts it~\cite{kwon2023vllm,gao2024cachedattention}. Switching to a cleaner region, a smaller model, or a more efficient GPU saves carbon, but the new deployment must recompute the history. Histories are often long: 40.3\% of queries in the ShareGPT dataset~\cite{sharegpt2023unsplit} carry at least 4K history tokens, and for them a cache miss more than triples the emissions of a query (Section~\ref{sec:motivation-miss}). The longer the conversation, the larger the carbon advantage a switch needs, a property we call \emph{context-induced carbon stickiness}. With 4K history tokens, for example, the new region must be 68\% cleaner than the current one (Section~\ref{sec:motivation-breakeven}). This trade-off is challenging. The lowest-carbon combination of model, GPU, and region changes from turn to turn as queries need models of different sizes and the cleanest region shifts. Whether a switch pays off depends on an uncertain future: how many turns remain, what they need, and how each placement changes the cache state and regional capacity that later queries of all sessions face.

Recent work has begun to address the environmental sustainability of LLM routing, that is, to reduce the carbon footprint of serving each query. Carbon-aware LLM routers choose a model and a region for every query~\cite{li2026faircarbon,sheshanarayana2026gar,more2026cedar}, but they treat the queries of a conversation as independent and therefore do not make this trade-off. This leaves a gap: no carbon-aware LLM router for interactive queries prices the carbon of the history recomputation that a deployment would incur. A deployment without the session's cache thus looks as cheap as if the history were free, and no router can decide online for each incoming query whether leaving the cache pays off in carbon.

We present CarbonRouter\footnote{We open-source CarbonRouter at \url{https://anonymous.4open.science/r/carbonrouter-artifact-00C6/}.}, an online routing framework that jointly selects the model, GPU type, and region for each query of a multi-turn conversation to minimize carbon emissions under quality and latency constraints. CarbonRouter combines the predicted quality of each model, the energy and latency profile of each model--GPU pair, the real-time carbon intensity of each region, and the KV-cache state of each deployment in a cache-aware carbon cost model, which charges every candidate deployment for recomputing the part of the history it does not cache. Its routing algorithm, \emph{Carbon- and Cache-Aware Routing} (CCAR), selects the feasible deployment with the lowest estimated carbon under this cost, so a session leaves its cache only when a cleaner region, a smaller model, or a more efficient GPU outweighs the history it rebuilds. A lightweight learned gate additionally keeps a session on its cached deployment when the proposed switch is likely to be reversed at the next turn. CarbonRouter sits at the request-routing layer of an LLM-based web service, between the web gateway and the serving backends in each region, and replaces only the routing policy of this layer. It is therefore easy to deploy: its inputs come from carbon intensity APIs, offline profiles, and the prefix caches that serving engines already keep, and the models, inference engines, and clients stay unchanged.

We evaluate CarbonRouter in trace-driven simulation on real conversations from ShareChat~\cite{yan2026sharechat} and ShareGPT~\cite{sharegpt2023unsplit}, with five LLMs from 1B to 70B parameters, three GPU types, and four regions whose real-world five-minute carbon intensity comes from Electricity Maps~\cite{electricitymaps2026api}. The baselines cover each class of prior work: session affinity with and without carbon-aware placement, and two of the latest carbon-aware LLM routers, GAR~\cite{sheshanarayana2026gar} and FairCarbon~\cite{li2026faircarbon}. CarbonRouter achieves the lowest carbon of all policies. Relative to FairCarbon, it reduces CO$_2$ by 16.1\% on average over the two datasets while increasing mean predicted quality, because it breaks the cache less often: CarbonRouter and FairCarbon both serve about 57\% of the ShareChat queries in the cleanest region, but CarbonRouter cuts the share of history tokens it recomputes from 27.7\% to 9.4\%. Consistent with carbon stickiness, the saving grows with the number of turns, to 23.1\% for sessions longer than 30 turns. 

Our contributions are as follows:
\begin{itemize}
\item We analyze grid carbon intensity and real conversations, and show that KV-cache locality changes carbon-aware routing for multi-turn LLM-based web services.
\item We propose CarbonRouter, a carbon- and cache-aware routing framework that jointly selects the model, GPU type, and region for each query, with a routing algorithm, CCAR, that prices history recomputation and avoids switches likely to be reversed.
\item We evaluate CarbonRouter with trace-driven experiments on real-world datasets against baselines that include the latest carbon-aware LLM routers, with analyses of conversation length, ablation, and sensitivity.
\end{itemize}

\section{Carbon and KV-Cache Analysis}
\label{sec:motivation}

\begin{figure}[t]
\centering
\includegraphics[width=\columnwidth]{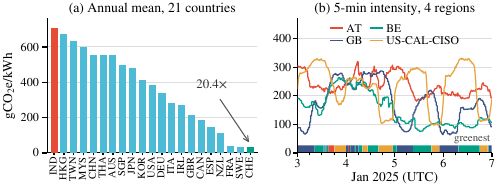}
\caption{Carbon intensity. (a) 2024 annual mean of 21 countries. (b) Five-minute intensity of our four regions (strip: cleanest region).}
\Description{Left: bar chart of annual carbon intensity for 21 countries, from India at about 705 gCO2e/kWh to Switzerland at about 35, a 20.4-fold spread. Right: four time series (AT, GB, BE, US-CAL-CISO) over four days varying between about 60 and 330 gCO2e/kWh, with a colored strip marking the greenest region.}
\label{fig:motiv-opportunity}
\end{figure}

\begin{figure}[t]
\centering
\includegraphics[width=\columnwidth]{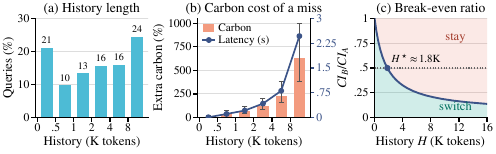}
\caption{Carbon cost of a KV-cache miss on ShareGPT. (a) History length. (b) Extra carbon of a miss (bars: median, error bars: interquartile range) and extra prefill latency (line). (c) Break-even ratio $CI_B/CI_A$ for leaving cached region $A$, with 128 new and 256 output tokens. $H^\star$: break-even history at a ratio of 0.5.}
\Description{Left: bar chart of query shares by history length, with 24 percent of queries above 8K tokens. Middle: extra carbon of a cache miss grows from about 3 percent for short histories to about 630 percent above 8K tokens, and extra latency grows to about 2.5 seconds. Right: the break-even ratio falls from 1 at zero history to 0.14 at 16K tokens; at a ratio of 0.5 the break-even history is about 1.8K tokens.}
\label{fig:motiv-cost}
\end{figure}

This section motivates CarbonRouter with two factors that a carbon-aware router for multi-turn LLM-based web services must weigh. The first is the grid carbon intensity of the regions it can route to. The second is the cost of recomputing a conversation's history when a query leaves its KV cache, and of how they interact. Our study seeks to answer three questions.
\begin{enumerate}
\item How much does grid carbon intensity differ across regions and change over time, and is the difference large enough to warrant routing LLM queries across regions?
\item How much does a KV-cache miss add to the carbon emissions of a query in real multi-turn conversations?
\item Given this cost, when does moving a session to a cleaner deployment reduce emissions?
\end{enumerate}

\subsection{Carbon Intensity across Regions and Time}
\label{sec:motivation-ci}

To quantify the opportunity, we use two carbon intensity datasets: the 2024 annual mean of 21 countries and areas from Ember~\cite{ember2026carbonintensity}, and the five-minute intensity of the four regions used in our evaluation, Austria (AT), Great Britain (GB), Belgium (BE), and California (US-CAL-CISO), in 2025 from Electricity Maps~\cite{electricitymaps2026api}.

Figure~\ref{fig:motiv-opportunity}(a) shows that annual grid carbon intensity spans a $20.4\times$ range, from 705~gCO$_2$e/kWh in India to 35~gCO$_2$e/kWh in Switzerland. The intensity also changes within hours. Figure~\ref{fig:motiv-opportunity}(b) shows a $2.7$--$5.9\times$ range within each of our four regions and a median $2.25\times$ gap between regions at the same time. No region is consistently the cleanest: each region is the lowest 13--32\% of the time.

\noindent\textbf{Key Takeaways.} \textit{Grid carbon intensity differs by an order of magnitude across regions and changes within hours, so routing LLM queries to low-carbon regions can reduce emissions. Because the cleanest region changes over time, the router must choose the region online, per query, using real-time carbon intensity.}

\subsection{Carbon Cost of a KV-Cache Miss}
\label{sec:motivation-miss}

A turn of a conversation is served with the KV cache of its history. This cache resides only on deployments that served earlier turns, and only until the backend evicts it; any other deployment must prefill the history again. To measure the carbon cost of this recomputation, we analyze random 2,000 ShareGPT sessions with the Mistral-7B--A100 profile and the default energy ratios of Section~\ref{sec:setup}. Because a query's emissions are its energy times the carbon intensity of its region, the relative increase from a miss on the same deployment does not depend on the carbon intensity.

Conversation histories are long. Figure~\ref{fig:motiv-cost}(a) shows that 40.3\% of queries carry at least 4K history tokens and 24.4\% at least 8K. Rebuilding such a history dominates the carbon cost of a query: a cache miss raises the emissions of a query by a median of 126\% for 2K--4K history tokens, 231\% for 4K--8K, and 632\% beyond 8K, and adds up to 2.5~s of prefill latency (Figure~\ref{fig:motiv-cost}(b)).

\noindent\textbf{Key Takeaways.} \textit{In multi-turn conversations, recomputing the history after a cache miss can emit several times more carbon than the query itself. A carbon-aware router that ignores this cost misjudges the carbon of every deployment that does not hold the session's cache, so it must charge recomputation on every cache miss.}

\subsection{When Does Switching Pay Off?}
\label{sec:motivation-breakeven}

Consider a session whose cache resides in region $A$ and whose current turn could move to a cleaner region $B$ with the same model and GPU. Staying in $A$ emits $CI_A E^{\mathrm{hit}}(H)$, where $E^{\mathrm{hit}}(H)$ is the energy of the current query, including its new input and output, when its history, the $H$ tokens of earlier queries and responses, is cached. Moving to $B$ emits $CI_B E^{\mathrm{miss}}(H)$, which adds the recomputation of the history. Moving thus reduces emissions only when
\begin{equation}
\frac{CI_B}{CI_A}<\tau(H)=\frac{E^{\mathrm{hit}}(H)}{E^{\mathrm{miss}}(H)}.
\label{eq:breakeven}
\end{equation}
The recomputation in $E^{\mathrm{miss}}(H)$ grows with the history, whereas a hit reuses the history at a small cost, so $\tau(H)$ falls as the conversation grows: the longer the conversation, the cleaner $B$ must be to justify a switch. We call this property \emph{context-induced carbon stickiness}. Under our default energy ratios, $B$ must be 68\% cleaner than $A$ with 4K history tokens (Figure~\ref{fig:motiv-cost}(c)). Note that this single-turn analysis shows the trend rather than the exact threshold.

\noindent\textbf{Key Takeaways.} \textit{The cleanest region is not necessarily the best choice for a conversation. The router must weigh each deployment's carbon intensity against the history recomputation it causes.}

\section{CarbonRouter Design}
\label{sec:design}

\subsection{Overview}
\label{sec:overview}

\begin{figure*}[t]
\centering
\includegraphics[width=\textwidth]{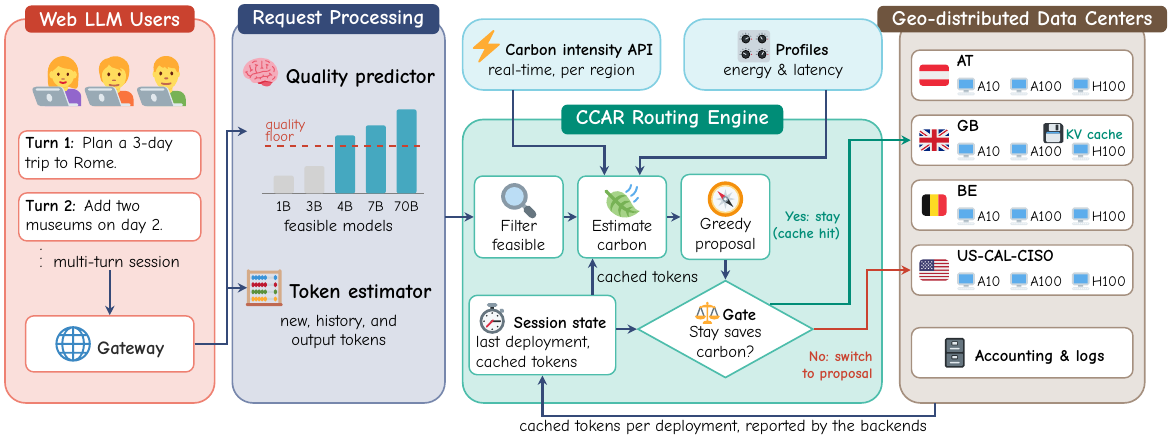}
\caption{Overview of CarbonRouter: each query goes to the lowest-carbon model, GPU, and region, given its cached history.}
\Description{Four colored panels from left to right. Web LLM Users: three users at laptops, two example turns of a chat session, and a gateway. Request Processing: a quality predictor with a bar chart of predicted quality for 1B to 70B models against a dashed quality floor, and a token estimator. Above the CCAR Routing Engine, two independent modules, a carbon intensity API and energy-and-latency profiles, feed its carbon estimate. CCAR Routing Engine: a pipeline of filter feasible, estimate carbon, and greedy proposal; a session-state box holding the last deployment and the cached tokens feeds the carbon estimate and a gate, drawn as a diamond, that asks whether staying saves carbon, with a green yes-branch staying on the cached deployment in GB and a red no-branch switching to the greedy proposal. Geo-distributed Data Centers: AT, GB, BE, and US-CAL-CISO, each with A10, A100, and H100 GPUs, GB holding the KV cache, plus accounting and logs. An arrow along the bottom carries the cached tokens per deployment, reported by the backends, to the session state.}
\label{fig:overview}
\end{figure*}

CarbonRouter is a carbon-aware and cache-aware online routing framework for multi-turn LLM-based web services. For each query, CarbonRouter jointly selects a model, a GPU type, and a region to minimize carbon emissions under quality and latency constraints. Figure~\ref{fig:overview} shows its architecture. When a query arrives, (i) the quality predictor estimates the response quality of each candidate model; (ii) the cost model estimates the energy, latency, and carbon emissions of every available deployment, using the energy and latency profile of each model--GPU pair and the current carbon intensity of each region from a carbon intensity API; (iii) the router selects the feasible deployment with the lowest estimated carbon, including the recomputation of uncached history, and a lightweight gate may keep the session on its cached deployment instead; and (iv) the session-state tracker records each turn's deployment and completion time and, at dispatch, obtains from the backends the number of history tokens cached on every candidate deployment. The profiles and the carbon intensity API are external inputs: an operator can plug in measured profiles or another carbon intensity provider.

\noindent\textbf{Deployability.} CarbonRouter can be deployed without changing the existing serving stack. It is a routing policy at the request-routing layer, between the web gateway and the serving backends in each region, so the models, inference engines, and clients stay as they are. Its inputs are already available: carbon intensity comes from public APIs~\cite{electricitymaps2026api}, energy and latency profiles are measured or estimated offline, and the cached history of each session is state that serving engines already keep for prefix reuse. The router need not query the serving engines for it on the critical path: the session-state tracker already records where and when each turn ran, so the router can track the history cached on each deployment locally from the TTL and from eviction notices that the engines send asynchronously, as cache-aware load balancers do~\cite{srivatsa2025preble,xia2026skywalker}. Because chat APIs commonly resend the full conversation with every request, the router also needs no new client state.

\subsection{System Model}
\label{sec:problem_formulation}

CarbonRouter serves a set $\mathcal S$ of multi-turn conversation sessions across geo-distributed data centers, with sets $\mathcal M$, $\mathcal G$, and $\mathcal R$ of models, GPU types, and regions. The $k$-th query $q_{s,k}$ of session $s$ arrives at time $t_{s,k}$ and carries the history of the earlier queries and responses. Turns of a session run sequentially, so the query is dispatched at $d_{s,k}=\max\{t_{s,k},f_{s,k-1}\}$, where $f_{s,k-1}$ is the completion time of the preceding turn ($d_{s,1}=t_{s,1}$). The router then selects a deployment
\begin{equation}
a_{s,k}=(m_{s,k},h_{s,k},r_{s,k})\in\mathcal{A}(d_{s,k})\subseteq\mathcal{M}\times\mathcal{G}\times\mathcal{R},
\label{eq:action}
\end{equation}
where $\mathcal A(t)$ is the set of deployments that the serving infrastructure makes available at time $t$ after resource and admission control.

KV caches are managed by the existing serving backend of each data center. The cache produced at a turn stays on the deployment that served it until its time-to-live $T^{\mathrm{TTL}}$ expires, measured from the turn's completion~\cite{liu2026droidspeak}, or until the backend evicts it in least-recently-used (LRU) order under memory pressure~\cite{kwon2023vllm,zheng2024sglang}. The backend reports the cache state to the router.

\subsection{Quality Prediction and Model Filtering}
\label{sec:quality-filter}

A quality predictor estimates the expected quality $Q_{s,k}(m)$ of each candidate model $m\in\mathcal M$ on a $[0,1]$ scale for the current query $q_{s,k}$. Any learned router predictor can serve this role. The quality requirement of a query is relative to its best candidate: with $Q^{\mathrm{tar}}_{s,k}=\max_{m\in\mathcal M}Q_{s,k}(m)$ and a required fraction $\beta_Q\in(0,1]$, the quality-feasible model set is
\begin{equation}
\mathcal{M}^{Q}_{s,k}(\beta_Q)=\left\{m\in\mathcal{M}:Q_{s,k}(m)\geq\beta_Q Q^{\mathrm{tar}}_{s,k}\right\}.
\label{eq:quality-set}
\end{equation}
The set is never empty, because the best candidate always qualifies, and it adapts to query difficulty: a small model is admitted only when it answers the query nearly as well as the best candidate. 

\subsection{Carbon Cost Model}
\label{sec:energy-carbon}

For candidate model $m$, let $N^{\mathrm{new}}_{s,k}(m)$ and $N^{\mathrm{hist}}_{s,k}(m)$ be the numbers of new input tokens and history tokens of query $q_{s,k}$ under $m$'s tokenizer, and $N^{\mathrm{out}}_{s,k}(m)$ its estimated output length. At dispatch, the backend returns for every candidate deployment $a=(m,h,r)$ the number $L_{s,k}(a)\le N^{\mathrm{hist}}_{s,k}(m)$ of history tokens cached there. A cache belongs to the model, GPU, and region that produced it, so $L_{s,k}(a)$ equals the full history on the deployment that served the previous turn while its cache is intact, is smaller after a partial eviction, and is zero where the session has no cache.

The profile of each model--GPU pair $(m,h)$ gives its energy coefficients $e^{\mathrm{in}}_{m,h}$, $e^{\mathrm{out}}_{m,h}$, and $e^{\mathrm{cache}}_{m,h}$, in kWh per token, for uncached input, output, and cached history tokens. Omitting $s,k$ and the model argument on the right, the estimated energy of candidate $a=(m,h,r)$ is
\begin{equation}
E_{s,k}(a)=e^{\mathrm{in}}_{m,h}\bigl(N^{\mathrm{new}}+N^{\mathrm{hist}}-L(a)\bigr)+e^{\mathrm{cache}}_{m,h}L(a)+e^{\mathrm{out}}_{m,h}N^{\mathrm{out}},
\label{eq:energy-estimate}
\end{equation}
which charges the cached part of the history at the cached-input rate and recomputes the rest. We write $E^{\mathrm{base}}_{s,k}(a)$ for this energy with the full history cached, $L_{s,k}(a)=N^{\mathrm{hist}}_{s,k}$; the difference $E_{s,k}(a)-E^{\mathrm{base}}_{s,k}(a)$ is the recomputation penalty of a cache miss.

With the regional carbon intensity $\mathrm{CI}_r(t)$ in gCO$_2$e/kWh reported by the carbon intensity API, the estimated emissions are
\begin{equation}
\widehat{C}_{s,k}(a)=\mathrm{CI}_r(d_{s,k})E_{s,k}(a),
\label{eq:carbon-estimate}
\end{equation}
using the intensity at dispatch for the whole request. We write $C_{s,k}(a_{s,k-1},a_{s,k})$ for the realized emissions of a served query, computed with the realized output length. We write $\gamma=e^{\mathrm{cache}}_{m,h}/e^{\mathrm{in}}_{m,h}$ for the cache discount and $\kappa=e^{\mathrm{out}}_{m,h}/e^{\mathrm{in}}_{m,h}$ for the output-to-input ratio.

\subsection{Latency Model}
\label{sec:latency-feasibility}

CarbonRouter estimates the time to first token (TTFT) from the profile's prefill throughput $v^{\mathrm{pre}}_{m,h}$ in tokens per second. Prefill covers only the uncached part of the history, so with the notation of Equation~\eqref{eq:energy-estimate}, the TTFT of candidate $a=(m,h,r)$ is
\begin{equation}
\mathrm{TTFT}_{s,k}(a)=\frac{N^{\mathrm{new}}+N^{\mathrm{hist}}-L(a)}{v^{\mathrm{pre}}_{m,h}}.
\label{eq:latency}
\end{equation}
It thus approximates TTFT by prefill execution time alone and excludes network transmission, queueing, scheduling, cache lookup, and first-token decoding, so it is a compute-side estimate rather than the end-to-end latency perceived by users. The model also neglects the variation of throughput with context length and concurrent load.

\subsection{Constraints and Optimization Goal}
\label{sec:constraints-goal}

Queries from different sessions arrive online and interleave. At $d_{s,k}$, the router knows only the information $\mathcal I_{s,k}$ available then, namely the observed contexts, previous actions and cache states, available deployments, and carbon intensities, but not future queries or how many turns remain. The online routing problem is
\begin{equation}
\begin{aligned}
\min_{\pi}\quad &\sum_{s\in\mathcal S}\sum_{k=1}^{K_s}C_{s,k}(a_{s,k-1},a_{s,k})\\
\text{s.t.}\quad &a_{s,k}=\pi(\mathcal I_{s,k})\in\mathcal A(d_{s,k}),\quad m_{s,k}\in\mathcal M^{Q}_{s,k}(\beta_Q),\quad\forall s,k,\\
&\mathrm{TTFT}_{s,k}(a_{s,k})\le D^{\mathrm{TTFT}},
\end{aligned}
\label{eq:global_objective}
\end{equation}
where $\pi$ is an online routing policy, $K_s$ is the number of turns of session $s$, and $D^{\mathrm{TTFT}}$ bounds the TTFT. We call a deployment that satisfies these constraints \emph{feasible}. If no deployment meets the TTFT bound but some meet every other constraint, typically because every deployment without the session's cache would have to recompute a long history, the router serves the query best-effort on the deployments with the shortest estimated TTFT rather than rejecting the turn.

\subsection{Routing Algorithm}
\label{sec:routing-algorithm}
\label{sec:advantage-gated-greedy}

Solving this online routing problem exactly is impractical for three reasons. First, the router must decide without knowing the future: whether keeping a session's cache pays off depends on how many turns remain, when they arrive, how carbon intensity evolves, and whether the backend keeps the cache, and none of this is known at dispatch. Second, decisions are coupled. Within a session, each action determines where the next turn finds the cache; across sessions, routing one session to a deployment uses capacity and KV memory that other sessions may need and can evict their caches. Third, the router sits on the critical path of every query, so its decision adds directly to the TTFT and must take negligible time; routers for web LLM services keep this overhead below a millisecond~\cite{li2026faircarbon}.

CarbonRouter therefore routes each query with \emph{Carbon- and Cache-Aware Routing} (CCAR), a fast per-query algorithm whose core is cache-aware selection: every candidate is charged the recomputation of the history it does not cache, so a session leaves its cache only when a cleaner region or a smaller model outweighs the rebuild. A lightweight learned gate covers the one effect this misses, the next turn.

\begin{algorithm}[t]
\caption{CCAR: Carbon- and Cache-Aware Routing}
\label{alg:advantage-gated-greedy}
\small
\begin{algorithmic}[1]
\REQUIRE Request $q_{s,k}$; previous deployment $b_{s,k-1}$ and completion time $f_{s,k-1}$; current information $\mathcal I_{s,k}$\rev{, including the cache state reported by the backend}; trained predictor $f_{\theta}$; threshold $\delta$
\ENSURE Routing action $a_{s,k}$, if a feasible deployment exists
\STATE Obtain $L_{s,k}(a)$ from the backend; find the feasible deployments (Equation~\eqref{eq:global_objective}), relaxing the TTFT bound if needed.
\IF{no deployment is feasible}
  \STATE Reject the query; \textbf{stop}.
\ENDIF
\FORALL{feasible $a$}
  \STATE Compute $\widehat C_{s,k}(a)$, including recomputation on a cache miss.
\ENDFOR
\STATE $g_{s,k}\gets\operatorname*{arg\,min}_{a}\widehat C_{s,k}(a)$; \ $a_{s,k}\gets g_{s,k}$.
\IF{$k>1$ \AND $g_{s,k}\ne b_{s,k-1}$ \AND $b_{s,k-1}$ feasible \AND $L_{s,k}(b_{s,k-1})>0$}
  \STATE Build features $z_{s,k}$; \ $\widehat A_{s,k}\gets f_{\theta}(z_{s,k})$.
  \IF{$\widehat A_{s,k}>\delta$}
    \STATE $a_{s,k}\gets b_{s,k-1}$.
  \ENDIF
\ENDIF
\STATE Execute $a_{s,k}$; record its deployment and completion time.
\end{algorithmic}
\end{algorithm}

As shown in Algorithm~\ref{alg:advantage-gated-greedy}, for each query $q_{s,k}$, CCAR first obtains the cached history $L_{s,k}(a)$ of every available deployment from the backend and keeps the feasible deployments, relaxing the TTFT bound as in Section~\ref{sec:constraints-goal} when needed; if none remains, the query is rejected. It then computes the estimated emissions $\widehat C_{s,k}(a)$ of every feasible deployment, including recomputation, and proposes the one with the lowest estimated carbon, $g_{s,k}\in\operatorname*{arg\,min}_{a}\widehat C_{s,k}(a)$. Finally, the gate decides whether to keep the session on its previous deployment instead, and the query runs on the chosen deployment.

The gate acts only at a \emph{switch opportunity}, where $g_{s,k}$ would leave a feasible deployment $b_{s,k-1}$ that still caches the session's history: if the session left and returned after the backend evicted that cache, the history would be recomputed again. A predictor $f_\theta$, an XGBoost regressor~\cite{chen2016xgboost}, estimates the carbon saved by staying over the current and next turn, $\widehat A_{s,k}=f_\theta(z_{s,k})$, from features available at dispatch, such as the estimated costs of both options and the cache state and memory pressure that the backend reports, and the router stays if $\widehat A_{s,k}>\delta$. The predictor is trained on other sessions to predict the realized two-turn difference, and $\delta$, including $+\infty$ (no gate), is selected by session-held-out validation.

CCAR runs in time linear in the number of available deployments, and the gate adds at most one predictor call. Like other per-query routers for web services~\cite{li2026faircarbon}, it thus trades global optimality for responsiveness and adds negligible latency to each turn.

\section{Evaluation}
\label{sec:evaluation}

\begin{table*}[t]
\centering
\caption{Overall performance, normalized to Session-Affinity $=1$. P95 TTFT is estimated from prefill time only (Equation~\eqref{eq:latency}). CO$_2$ saving: CarbonRouter's saving over each row, with 95\% confidence intervals. $^\dagger$Adapted to our setting.}
\label{tab:overall}
\small
\setlength{\tabcolsep}{4pt}
\begin{tabular}{lrrrcrrrcc}
\toprule
 & \multicolumn{4}{c}{ShareChat} & \multicolumn{4}{c}{ShareGPT} & Average\\
\cmidrule(lr){2-5}\cmidrule(lr){6-9}\cmidrule(lr){10-10}
Method & CO$_2\downarrow$ & Quality$\uparrow$ & P95 TTFT (s)$\downarrow$ & CO$_2$ saving (\%) & CO$_2\downarrow$ & Quality$\uparrow$ & P95 TTFT (s)$\downarrow$ & CO$_2$ saving (\%) & CO$_2$ saving (\%)\\
\midrule
Session-Affinity & 1.000 & 1.000 & 0.05 & 53.7 [52.3, 55.1] & 1.000 & 1.000 & 0.04 & 54.9 [52.8, 56.9] & 54.3 [53.1, 55.6]\\
Carbon-Affinity & 0.586 & 0.999 & 0.09 & 21.0 [20.0, 22.0] & 0.575 & 1.000 & 0.05 & 21.6 [19.7, 23.5] & 21.3 [20.3, 22.4]\\
GAR$^\dagger$ & 0.630 & 0.973 & 1.72 & 26.5 [24.2, 28.8] & 0.660 & 0.976 & 0.56 & 31.6 [28.5, 34.8] & 29.1 [27.1, 30.9]\\
FairCarbon & 0.545 & 0.973 & 1.67 & 15.1 [13.4, 16.9] & 0.544 & 0.976 & 0.53 & 17.0 [14.3, 19.7] & 16.1 [14.4, 17.6]\\
\textbf{CarbonRouter (ours)} & \textbf{0.463} & \textbf{0.979} & 0.52 & -- & \textbf{0.451} & \textbf{0.979} & 0.20 & -- & --\\
\bottomrule
\end{tabular}
\end{table*}

We evaluate our proposed routing framework CarbonRouter, which combines carbon-aware and cache-aware routing, using real-world workloads and trace-driven simulation with an analytical energy and latency model. 

\subsection{Experimental Setup}
\label{sec:setup}

\noindent\textbf{Workloads and arrivals.} 
We use two real-world LLM conversation datasets, ShareChat~\cite{yan2026sharechat} and ShareGPT~\cite{sharegpt2023unsplit}, to evaluate our CarbonRouter framework. The queries follow the multi-turn chat arrival model of UniCache~\cite{ouyang2026unicache}. A query that arrives before its predecessor completes waits for it. All policies share the same arrivals.

\noindent\textbf{Models and GPUs.} Five instruction-tuned models (Gemma-3-1B, Llama-3.2-3B, Gemma-3-4B, Mistral-7B-v0.2, and Llama-3.1-70B) run on A10, A100-40GB, and H100-SXM in each of four regions, giving 60 deployments. Each model runs on one GPU except Llama-3.1-70B, which uses eight A10, eight A100-40GB, or four H100 GPUs. These counts divide its attention heads, as tensor parallelism requires, and leave 32--147~GB for the KV cache after its 141~GB of 16-bit weights (Appendix~\ref{sec:latency-profiles}). Quality predictions come from the released R2-Router predictors~\cite{xue2026r2router,r2router2026release}, which score every model once per query from the current query and recent context.

\noindent\textbf{Energy and latency profiles.} The energy and latency profiles are a pluggable component of CarbonRouter (Figure~\ref{fig:overview}), so an operator can load measured profiles of its own deployments. In our evaluation, they come from an analytical model built from published data rather than from measurements (Appendix~\ref{sec:impl-params}). Scaling all energy coefficients by a common factor does not change which candidate has the lowest estimated carbon, so routing depends only on the relative energy of deployments and on the ratios $\kappa$ and $\gamma$. We set $\kappa=6$ and $\gamma=0.05$ from API price ratios~\cite{openai2026pricing,anthropic2026pricing,google2026pricing}, following AgentDecarbonizer~\cite{yan2026agentdecarbonizer}. Measurements agree: a cached input token costs 0.045--0.084 of an uncached one~\cite{cho2026librarian}, and output-to-input energy ratios are about 3--8 under batched serving~\cite{wu2026energylaw,caravaca2025prompts}.

\noindent\textbf{Carbon intensity signals.} We use five-minute life-cycle carbon intensity from Electricity Maps~\cite{electricitymaps2026api} for AT, GB, BE, and US-CAL-CISO in 2025. Each query uses the latest value at or before its dispatch time.

\noindent\textbf{Constraints and cache.} We use $\beta_Q=0.85$, so each query admits the models whose predicted quality is at least 85\% of its best candidate's, a TTFT bound of 5~s, enforced best-effort as in Section~\ref{sec:constraints-goal}, and each model's context limit. A KV cache stays on its deployment for at most 30 minutes after its turn completes, unless the emulated backend evicts it earlier (below).

\noindent\textbf{Emulated backend.} Without real geo-distributed data centers, we cannot observe what their backends would report, so we emulate availability and caching in the replay, as prior trace-driven studies of geographical load balancing emulate capacity~\cite{liu2011greening,sukprasert2024limitations,li2024equitable,bernhard2026cleanestgrid}. Each region holds at most $K=c\,P$ sessions, where $P$ is the peak number of simultaneously active sessions, and a deployment is available only if its region already holds the session or has a free slot. The KV memory of a region holds $K\bar C$ tokens for the mean query context $\bar C$; when a new cache does not fit, a deployment evicts idle caches in LRU order, starting with those of sessions that have left~\cite{kwon2023vllm,zheng2024sglang}, and reports the remaining cached history $L_{s,k}(a)$ to the router. We use $c=0.5$, a total capacity of twice the peak demand, and vary $c$ in Section~\ref{sec:sensitivity}. Each region’s capacity is split equally across its three GPU types

\noindent\textbf{Baselines.} We compare CarbonRouter against four baselines. (i) \emph{Session-Affinity} keeps each session on one deployment while it remains feasible; it is cache-aware but carbon-agnostic and serves as the reference for all ratios. (ii) \emph{Carbon-Affinity} chooses the lowest-carbon feasible deployment when a session starts and keeps the session there while its cache remains valid and the deployment remains feasible, so it considers carbon only when it places a session. (iii) \emph{GAR$^\dagger$} adapts GAR~\cite{sheshanarayana2026gar}, a recent carbon-aware LLM router, to choose the lowest-carbon model and GPU within one fixed region; it is carbon-aware but cache-oblivious. (iv) \emph{FairCarbon}~\cite{li2026faircarbon}, a recent carbon-aware LLM router for web services, chooses the lowest-carbon deployment across all regions. We omit its fairness term, which balances model capacity across user groups and requires group labels that our traces lack. 

\noindent\textbf{Metrics.} We report (i) total operational CO$_2$, (ii) the mean predicted quality of the selected models, and (iii) the 95th-percentile TTFT estimated by the latency model of Section~\ref{sec:latency-feasibility}.

\subsection{Overall Performance}
\label{sec:overall}

Table~\ref{tab:overall} shows the overall performance of CarbonRouter compared to other baselines. CarbonRouter achieves the lowest CO$_2$ of all methods on both datasets. Overall, it reduces CO$_2$ by 16.1--54.3\% on average relative to the four baselines. More specifically, CarbonRouter reduces CO$_2$ by 54.3\% relative to Session-Affinity, by 21.3\% relative to Carbon-Affinity, by 29.1\% relative to GAR$^\dagger$, and by 16.1\% relative to the SOTA baseline FairCarbon.

\noindent\textbf{Where the saving comes from.} GAR$^\dagger$ and FairCarbon both ignore the cache, but FairCarbon emits 13.4\% less CO$_2$ than GAR$^\dagger$, which isolates the benefit of routing across regions. Both move sessions between small and large models from turn to turn and recompute more than a quarter of the history tokens (28.6\% and 27.7\% on ShareChat). CarbonRouter avoids most of this recomputation. It serves the same share of queries in the cleanest region as FairCarbon (57\% on ShareChat and 67\% on ShareGPT), so its saving over FairCarbon comes from breaking the cache less often: it switches deployments 5.58 times per ShareChat session instead of 7.14, and the share of history tokens it recomputes falls from 27.7\% to 9.4\% (from 25.5\% to 10.0\% on ShareGPT). Carbon-Affinity fails in the opposite way. It serves as many queries in the cleanest region as FairCarbon (57.5\% on ShareChat) and recomputes only 2.9\% of the history tokens, but a session that moves to Llama-3.1-70B after a hard query stays there, so 87.5\% of its ShareChat queries run on the 70B model and it emits more CO$_2$ than FairCarbon on both datasets. CarbonRouter avoids both failures because it leaves the cache only when a cleaner region or a smaller model outweighs the recomputation.

\noindent\textbf{Quality.} Session-Affinity has the highest quality, but as a side effect of over-provisioning: it re-binds almost only when the bound model fails the quality requirement, so a session moves to a larger model after a hard query and rarely returns, and 90.0\% of ShareChat queries (82.0\% on ShareGPT) run on Llama-3.1-70B. Carbon-Affinity reaches almost the same quality for the same reason. Compared with Session-Affinity, CarbonRouter lowers mean predicted quality by only 2.1\% while reducing CO$_2$ by 54.3\% on average. CarbonRouter's mean predicted quality also exceeds FairCarbon's on both datasets, because pricing the cache keeps a session on its cached large model instead of moving it to a small one for a single easy turn.

\noindent\textbf{Latency.} Because it recomputes less history, CarbonRouter also has a lower estimated TTFT than the cache-oblivious routers. Its estimated P95 TTFT is 0.52~s on ShareChat, against 1.72~s for GAR$^\dagger$ and 1.67~s for FairCarbon (0.20~s against 0.56~s and 0.53~s on ShareGPT), and only 0.18\% of its ShareChat queries are served best-effort above the 5~s bound, against 0.76\% for FairCarbon. Session-Affinity and Carbon-Affinity have the lowest TTFT because they rarely leave their cache, at 2.16 and 1.27 times the CO$_2$ of CarbonRouter on ShareChat.

\subsection{Effect of Session Turns}
\label{sec:depth}

\begin{figure*}[t]
\centering
\includegraphics[width=\textwidth]{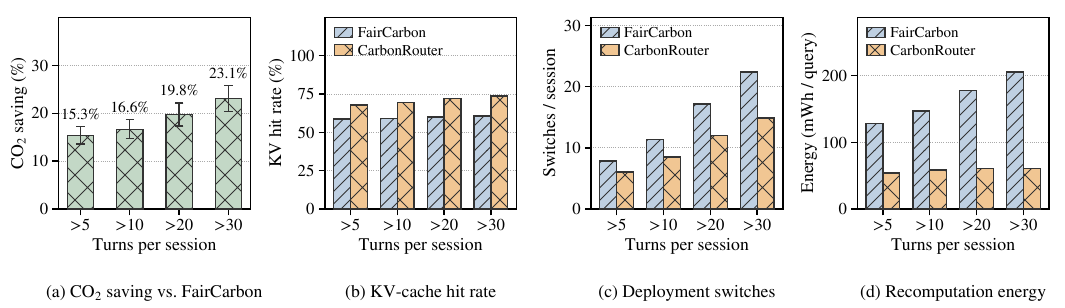}
\caption{CarbonRouter versus FairCarbon on ShareChat by session length, with 95\% confidence intervals in (a).}
\Description{Four bar-chart panels over four nested session-length groups (more than 5, 10, 20, and 30 turns) on ShareChat. CarbonRouter's carbon saving over FairCarbon is 15.3, 16.6, 19.8, and 23.1 percent. FairCarbon's cache hit rate stays near 59 to 61 percent, whereas CarbonRouter's rises from 67.9 to 73.7 percent. FairCarbon's switches per session rise from 7.8 to 22.4 and its recomputation energy per query from 129 to 205 mWh, while CarbonRouter's switches rise from 6.1 to 14.9 and its recomputation energy stays between 54 and 61 mWh.}
\label{fig:depth}
\end{figure*}

Figure~\ref{fig:depth} compares FairCarbon and CarbonRouter on ShareChat for sessions longer than 5, 10, 20, and 30 turns, using the same setting and sessions as Table~\ref{tab:overall}. Each group includes all sessions above its threshold. ShareGPT shows the same trends with fewer sessions of many turns.

\noindent\textbf{Savings grow with the number of turns.} CarbonRouter's CO$_2$ saving over FairCarbon rises from \rev{15.3\%} for sessions longer than 5 turns to \rev{16.6\%}, \rev{19.8\%}, and \rev{23.1\%} for sessions longer than 10, 20, and 30 turns (Figure~\ref{fig:depth}(a)). This is context-induced carbon stickiness (Section~\ref{sec:motivation-breakeven}) at work: the longer the history, the more a cache-breaking switch costs.

\noindent\textbf{Mechanism.} FairCarbon decides each query independently, so its cache reuse does not improve as sessions have more turns: its hit rate stays between \rev{58.7\% and 60.7\%} (Figure~\ref{fig:depth}(b)). Its switches per session grow from \rev{7.8 to 22.4} (Figure~\ref{fig:depth}(c)), and because each switch rebuilds a longer history in sessions with more turns, its recomputation energy per query grows from \rev{129 to 205}~mWh (Figure~\ref{fig:depth}(d)). CarbonRouter prices these rebuilds, and its gate blocks the switches whose next-turn cost outweighs their current saving. For sessions longer than 30 turns, it raises the hit rate from \rev{60.7\%} to \rev{73.7\%}, reduces switches per session by \rev{33\%} (from \rev{22.4 to 14.9}), and cuts recomputation energy per query by \rev{70\%} (from \rev{205 to 61}~mWh). CarbonRouter's recomputation energy per query stays between \rev{54 and 61}~mWh across all groups, even though a miss costs more in sessions with more turns: \rev{the share of its turns that miss at least part of the cache falls from 32.1\% to 26.3\%, and 58.8\% of these misses in sessions longer than 30 turns occur in the first half of the session, while the history is still short, against 50.0\% for FairCarbon.}

\subsection{Ablation Study}
\label{sec:ablation}

\begin{table}[t]
\centering
\caption{Ablation on ShareChat, normalized to Session-Affinity $=1$. TTFT: estimated P95, as in Table~\ref{tab:overall}. $\Delta$: increase over CarbonRouter, with 95\% confidence intervals.}
\label{tab:ablation}
\footnotesize
\setlength{\tabcolsep}{3pt}
\begin{tabular}{lrrcrr}
\toprule
Variant & CO$_2\downarrow$ & Energy$\downarrow$ & TTFT (s)$\downarrow$ & $\Delta$CO$_2$ (\%) & $\Delta$TTFT (\%)\\
\midrule
\textbf{CarbonRouter} & \textbf{0.463} & \textbf{0.826} & \textbf{0.52} & -- & --\\
w/o Gate & 0.479 & 0.854 & 0.89 & 3.5 [2.6, 4.6] & 71.4 [60.3, 81.9]\\
w/o Pricing & 0.504 & 0.907 & 0.74 & 8.9 [7.8, 10.2] & 42.4 [35.7, 49.6]\\
\bottomrule
\end{tabular}
\end{table}

\begin{figure}[t]
\centering
\includegraphics[width=\columnwidth]{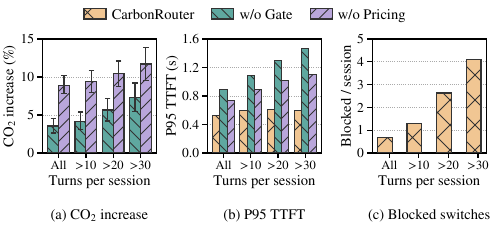}
\caption{Ablation on ShareChat by session length. (a) CO$_2$ increase over the full model, with 95\% confidence intervals. (b) 95th-percentile estimated TTFT. (c) Switches blocked by the gate per session.}
\Description{Three bar-chart panels over four session-length groups (all sessions and sessions longer than 10, 20, and 30 turns) on ShareChat. (a) The CO2 increase without the gate rises from 3.5 to 7.3 percent, and without pricing from 8.9 to 11.7 percent. (b) CarbonRouter's estimated P95 TTFT stays between 0.52 and 0.61 seconds, against 0.89 to 1.47 seconds without the gate and 0.74 to 1.10 seconds without pricing. (c) The gate blocks 0.7 switches per session over all sessions and 4.1 for sessions longer than 30 turns.}
\label{fig:ablation-turns}
\end{figure}

To quantify the contribution of cache pricing and the gate, we evaluate two variants of CarbonRouter in the setting of Table~\ref{tab:overall}. \emph{w/o Gate} disables the gate but keeps cache pricing. \emph{w/o Pricing} keeps the gate but drops the recomputation term from the carbon estimate, so the proposal ignores the cache, and retrains the gate on this proposal. Table~\ref{tab:ablation} reports the results.

\noindent\textbf{Cache awareness.} Both cache mechanisms reduce carbon and latency. Without pricing, CO$_2$ rises by 8.9\%. The proposal no longer charges recomputation and proposes many switches whose recomputation exceeds their saving. The gate can only veto them, so it also blocks switches to smaller models that would save carbon, and more queries stay on the large model, which raises energy (Table~\ref{tab:ablation}). Removing the gate raises CO$_2$ by only 3.5\% but the estimated P95 TTFT by 71.4\%, much more than removing pricing does (Table~\ref{tab:ablation}), because a switch reversed at the next turn recomputes the history twice.

\noindent\textbf{Session length.} The gate becomes increasingly important as conversations grow longer. Removing the gate raises CO$_2$ by 3.5\% over all sessions but by 7.3\% for sessions longer than 30 turns, twice as much (Figure~\ref{fig:ablation-turns}(a)), whereas the effect of removing pricing grows only from 8.9\% to 11.7\%. In a long session, a reversed switch recomputes a long history twice, so the gate intervenes more often and saves more: it blocks 4.1 switches per session longer than 30 turns, against 0.7 over all sessions (Figure~\ref{fig:ablation-turns}(c)), and keeps the estimated P95 TTFT between 0.52~s and 0.61~s in every group, whereas without the gate it grows from 0.89~s to 1.47~s for sessions longer than 30 turns and exceeds that without pricing in every group (Figure~\ref{fig:ablation-turns}(b)).

Overall, both components are necessary. Pricing contributes more to the carbon saving and the gate more to the latency: pricing prevents switches whose recomputation exceeds their saving, and the gate prevents switches that the next turn would reverse.

\begingroup
\subsection{Sensitivity Analysis}
\label{sec:sensitivity}

\begin{table}[t]
\centering
\caption{CO$_2$ saving (\%) of CarbonRouter over FairCarbon on ShareChat by $\gamma$ and $\kappa$, with 95\% confidence intervals.}
\label{tab:sensitivity}
\small
\setlength{\tabcolsep}{3pt}
\begin{tabular}{lccc}
\toprule
$\gamma$ & $\kappa=5$ & $\kappa=6$ & $\kappa=8$\\
\midrule
0.025 & 18.8 [16.7, 21.0] & 16.7 [14.7, 18.8] & 13.6 [11.9, 15.4]\\
0.05 & 16.8 [14.9, 18.7] & 15.1 [13.4, 16.9] & 12.6 [11.1, 14.1]\\
0.1 & 12.9 [11.2, 14.7] & 12.7 [11.1, 14.1] & 9.7 [8.4, 11.0]\\
\bottomrule
\end{tabular}
\end{table}

\begin{table}[t]
\centering
\caption{CO$_2$ saving (\%) of CarbonRouter over FairCarbon on ShareChat by $c$ and $\beta_Q$, with 95\% confidence intervals.}
\label{tab:sensitivity-capacity}
\small
\setlength{\tabcolsep}{3pt}
\begin{tabular}{lccc}
\toprule
$c$ & $\beta_Q=0.8$ & $\beta_Q=0.85$ & $\beta_Q=0.9$\\
\midrule
0.5 & 17.8 [16.1, 19.4] & 15.1 [13.4, 16.9] & 13.2 [11.9, 14.6]\\
0.35 & 11.4 [9.7, 13.1] & 11.2 [9.4, 12.8] & 11.4 [9.7, 13.2]\\
0.25 & 10.7 [9.0, 12.5] & 9.4 [7.7, 11.0] & 10.5 [8.6, 12.3]\\
\bottomrule
\end{tabular}
\end{table}

\rev{Two energy ratios, the cache discount $\gamma$ and the output-to-input ratio $\kappa$, set how much a cache miss costs (Section~\ref{sec:setup}). The regional capacity $c$, which is emulated, and the quality requirement $\beta_Q$, which is set by the operator, determine which deployments a query can use. We therefore vary these four parameters to test the robustness of our results.} Overall, CarbonRouter saves 9.4--18.8\% over FairCarbon in every setting we test (Tables~\ref{tab:sensitivity} and~\ref{tab:sensitivity-capacity}), and the saving is largest when a cache miss costs much more than a hit and when the cleanest region has spare capacity.

\noindent\textbf{Energy ratios.} Current LLM APIs price a cached input token at 2.5--10\% of a regular input token and an output token at 5--8 times an input token~\cite{openai2026pricing,anthropic2026pricing,google2026pricing}. We therefore vary $\gamma\in\{0.025,0.05,0.1\}$ and $\kappa\in\{5,6,8\}$ to cover both ranges. CarbonRouter saves \rev{9.7--18.8\%} over FairCarbon in all nine settings (Table~\ref{tab:sensitivity}). The saving is largest at $\gamma=0.025$, $\kappa=5$ and smallest at $\gamma=0.1$, $\kappa=8$, following the break-even analysis of Section~\ref{sec:motivation-breakeven}: a smaller $\gamma$ makes recomputation relatively more expensive, and a smaller $\kappa$ makes the history a larger share of a query's energy, so both favor keeping a session on its cache.

\rev{\noindent\textbf{Regional capacity and quality requirement.} We vary $c\in\{0.5,\allowbreak 0.35,\allowbreak 0.25\}$, from a total capacity of twice the peak demand to one equal to it, with the KV memory of each region scaled accordingly, and $\beta_Q\in\{0.8,\allowbreak 0.85,\allowbreak 0.9\}$. Tighter capacity reduces the saving (Table~\ref{tab:sensitivity-capacity}): at $\beta_Q=0.85$, it falls from 15.1\% at $c=0.5$ to 9.4\% at $c=0.25$, because the cleanest region fills earlier and both policies serve more queries in dirtier regions. The quality requirement matters only when capacity is ample: at $c=0.5$, the saving falls from 17.8\% to 13.2\% as $\beta_Q$ rises from 0.8 to 0.9, because fewer models qualify and FairCarbon switches less, leaving less recomputation to avoid. At $c=0.35$ and $0.25$, the saving stays within 11.2--11.4\% and 9.4--10.7\%.}

\endgroup

\section{Related Work}
\label{sec:related}

Carbon-aware computing cuts operational carbon by exploiting the temporal and spatial variation of grid carbon intensity. Temporal shifting delays flexible work to cleaner hours~\cite{radovanovic2022carbonaware,wiesner2021letswait} or scales it with the carbon signal~\cite{hanafy2024carbonscaler}, and spatial shifting moves load across data centers through geographical load balancing~\cite{liu2011greening,li2024equitable}, whose benefit simple policies largely capture~\cite{sukprasert2024limitations}. These systems treat the workload as fixed and assume that it is deferrable or stateless, so moving it costs little.

\begin{table}[t]
\centering
\caption{Comparison with existing carbon-aware methods.}
\label{tab:related}
\small
\setlength{\tabcolsep}{3pt}
\begin{tabular}{@{}lccccc@{}}
\toprule
Method & Model & GPU & Region & \twoline{KV\\cache} & \twoline{Interactive\\workload}\\
\midrule
\multicolumn{6}{@{}l}{\emph{Carbon-aware computing}}\\
Temporal shifting~\cite{radovanovic2022carbonaware,wiesner2021letswait,hanafy2024carbonscaler} & \xmark & \xmark & \xmark & \xmark & \xmark\\
Geo.\ load balancing~\cite{liu2011greening,li2024equitable} & \xmark & \xmark & \cmark & \xmark & \cmark\\
\multicolumn{6}{@{}l}{\emph{Carbon-aware LLM inference}}\\
Bernhard et al.~\cite{bernhard2026cleanestgrid} & \xmark & \xmark & \cmark & \xmark & \cmark\\
AgentDecarbonizer~\cite{yan2026agentdecarbonizer} & \xmark & \xmark & \cmark & \cmark & \xmark\\
\multicolumn{6}{@{}l}{\emph{Carbon-aware LLM routing}}\\
GAR~\cite{sheshanarayana2026gar} & \cmark & \xmark & \xmark & \xmark & \cmark\\
CEDAR~\cite{more2026cedar} & \cmark & \xmark & \cmark & \xmark & \xmark\\
FairCarbon~\cite{li2026faircarbon} & \cmark & \cmark & \cmark & \xmark & \cmark\\
\midrule
\textbf{CarbonRouter} & \cmark & \cmark & \cmark & \cmark & \cmark\\
\bottomrule
\end{tabular}
\end{table}

Recent work brings these ideas to LLM inference but keeps the model fixed. Bernhard and Yardimci~\cite{bernhard2026cleanestgrid} send each query to the currently cleanest region. AgentDecarbonizer~\cite{yan2026agentdecarbonizer} migrates agent tasks between grids and charges a recomputation penalty for the lost cache, but it plans each task offline from its estimated duration, whereas a chat turn must be served on arrival and the length of a conversation is unknown. GreenCache~\cite{tian2025greencache} sizes the KV cache of one deployment by carbon but does not move requests. Independently of carbon, cache-aware LLM serving systems schedule a query on the instance that holds its KV cache, within a cluster~\cite{srivatsa2025preble} or across regions~\cite{xia2026skywalker,toniolo2026gorgo}, to reduce latency.

LLM routing changes the computation itself by choosing a model for each query, trading answer quality against cost~\cite{chen2023frugalgpt,ding2024hybridllm,ong2025routellm,xue2026r2router}. Carbon-aware LLM routers make this choice by carbon. GAR~\cite{sheshanarayana2026gar} picks the lowest-carbon model under accuracy and latency constraints, CEDAR~\cite{more2026cedar} routes agentic requests across regions and models, and FairCarbon~\cite{li2026faircarbon} jointly selects the model, GPU, and region while balancing carbon, quality, and regional fairness. These routers treat the queries of a conversation as independent and do not price the recomputation caused by leaving a cache. CarbonRouter additionally prices the cache miss of every candidate deployment (Table~\ref{tab:related}). We use GAR and FairCarbon as baselines (Section~\ref{sec:setup}) but not CEDAR, which in our setting, without deferral or monetary cost, makes the same cache-oblivious choice per request as FairCarbon over a smaller action space.

\section{Limitations}
\label{sec:limitations}

Our evaluation is trace-driven simulation: energy and prefill latency come from an analytical model built from published data rather than from measurements on our own hardware, and answer quality is predicted rather than judged. CarbonRouter's decisions depend only on the ratios between energy coefficients, and its saving persists across the ranges we test, but validating the cost model against measured hit and miss energy on production serving engines remains future work, as does modeling network latency, queueing, and cross-region KV-cache transfer.

\section{Conclusion}
\label{sec:conclusion}

In this paper, we studied carbon-aware routing of multi-turn conversations in LLM-based web services, where moving a turn to a cleaner region or a smaller model forces its history to be recomputed. We showed that a cache miss can more than triple the emissions of a query with a long history. We developed CarbonRouter, which prices history recomputation for every candidate deployment and uses a lightweight gate to avoid switches likely to be reversed. On two real-world datasets, it reduces CO$_2$ by 16.1\% on average relative to FairCarbon and by 23.1\% for sessions longer than 30 turns, while improving predicted quality and estimated latency.

\bibliographystyle{ACM-Reference-Format}
\bibliography{ref}


\begin{thebibliography}{43}


\ifx \showCODEN    \undefined \def \showCODEN     #1{\unskip}     \fi
\ifx \showISBNx    \undefined \def \showISBNx     #1{\unskip}     \fi
\ifx \showISBNxiii \undefined \def \showISBNxiii  #1{\unskip}     \fi
\ifx \showISSN     \undefined \def \showISSN      #1{\unskip}     \fi
\ifx \showLCCN     \undefined \def \showLCCN      #1{\unskip}     \fi
\ifx \shownote     \undefined \def \shownote      #1{#1}          \fi
\ifx \showarticletitle \undefined \def \showarticletitle #1{#1}   \fi
\ifx \showURL      \undefined \def \showURL       {\relax}        \fi
\providecommand\bibfield[2]{#2}
\providecommand\bibinfo[2]{#2}
\providecommand\natexlab[1]{#1}
\providecommand\showeprint[2][]{arXiv:#2}
\makeatletter
\@ifundefined{NAT@parse@date}{}{\let\NAT@parse@date@orig\NAT@parse@date}
\@ifundefined{NAT@parse@date}{}{\def\NAT@parse@date#1#2#3#4#5#6@@{\NAT@parse@date@orig#1#2#3#4#5#6@@\def\NAT@tempyear{0000}\def\NAT@tempexlab{{?}}\ifx\NAT@year\NAT@tempyear\ifx\NAT@exlab\NAT@tempexlab\def\NAT@date{[n.\,d.]}\else\edef\NAT@date{[n.\,d.]\NAT@exlab}\fi\fi}}
\makeatother

\bibitem[{anon8231489123}(2023)]%
        {sharegpt2023unsplit}
\bibfield{author}{\bibinfo{person}{{anon8231489123}}.}
  \bibinfo{year}{2023}\natexlab{}.
\newblock \bibinfo{title}{{ShareGPT\_Vicuna\_unfiltered}: {HTML}-Cleaned Raw
  Conversations}.
\newblock \bibinfo{howpublished}{Hugging Face Datasets}.
\newblock
\urldef\tempurl%
\url{https://huggingface.co/datasets/anon8231489123/ShareGPT_Vicuna_unfiltered}
\showURL{%
\tempurl}
\newblock
\shownote{Unsplit HTML-cleaned raw files, parts 1 and 2; revision 192ab2185289;
  accessed September 14, 2026}.


\bibitem[Antepara et~al\mbox{.}(2025)]%
        {antepara2025energy}
\bibfield{author}{\bibinfo{person}{Oscar Antepara}, \bibinfo{person}{Zhengji
  Zhao}, \bibinfo{person}{Brian Austin}, \bibinfo{person}{Nan Ding},
  \bibinfo{person}{Leonid Oliker}, \bibinfo{person}{Nicholas~J. Wright}, {and}
  \bibinfo{person}{Samuel Williams}.} \bibinfo{year}{2025}\natexlab{}.
\newblock \showarticletitle{Benchmark-driven Models for Energy Analysis and
  Attribution of {GPU}-Accelerated Supercomputing}. In
  \bibinfo{booktitle}{\emph{Proceedings of the International Conference for
  High Performance Computing, Networking, Storage and Analysis (SC '25)}}.
  \bibinfo{address}{St Louis, MO, USA}.
\newblock
\href{https://doi.org/10.1145/3712285.3759815}{doi:\nolinkurl{10.1145/3712285.3759815}}


\bibitem[{Anthropic}(2026)]%
        {anthropic2026pricing}
\bibfield{author}{\bibinfo{person}{{Anthropic}}.}
  \bibinfo{year}{2026}\natexlab{}.
\newblock \bibinfo{title}{Claude {API} Pricing}.
\newblock
  \bibinfo{howpublished}{\url{https://platform.claude.com/docs/en/about-claude/pricing}}.
\newblock
\newblock
\shownote{Accessed: 2026-10-16}.


\bibitem[Bernhard and Yardimci(2026)]%
        {bernhard2026cleanestgrid}
\bibfield{author}{\bibinfo{person}{Aleks Bernhard} {and}
  \bibinfo{person}{Arif~Baran Yardimci}.} \bibinfo{year}{2026}\natexlab{}.
\newblock \bibinfo{title}{Routing {LLM} Inference to the Cleanest Grid in Real
  Time}.
\newblock
\showeprint[arxiv]{2608.06188}
\urldef\tempurl%
\url{https://arxiv.org/abs/2608.06188}
\showURL{%
\tempurl}


\bibitem[Caravaca et~al\mbox{.}(2025)]%
        {caravaca2025prompts}
\bibfield{author}{\bibinfo{person}{Francisco Caravaca},
  \bibinfo{person}{{\'A}ngel Cuevas}, {and} \bibinfo{person}{Rub{\'e}n
  Cuevas}.} \bibinfo{year}{2025}\natexlab{}.
\newblock \bibinfo{title}{From Prompts to Power: Measuring the Energy Footprint
  of {LLM} Inference}.
\newblock
\showeprint[arxiv]{2511.05597}~[cs.AI]
\href{https://doi.org/10.48550/arXiv.2511.05597}{doi:\nolinkurl{10.48550/arXiv.2511.05597}}


\bibitem[Chen et~al\mbox{.}(2024)]%
        {chen2023frugalgpt}
\bibfield{author}{\bibinfo{person}{Lingjiao Chen}, \bibinfo{person}{Matei
  Zaharia}, {and} \bibinfo{person}{James Zou}.}
  \bibinfo{year}{2024}\natexlab{}.
\newblock \showarticletitle{{FrugalGPT}: How to Use Large Language Models While
  Reducing Cost and Improving Performance}.
\newblock \bibinfo{journal}{\emph{Transactions on Machine Learning Research}}
  (\bibinfo{year}{2024}).
\newblock
\urldef\tempurl%
\url{https://openreview.net/forum?id=cSimKw5p6R}
\showURL{%
\tempurl}


\bibitem[Chen and Guestrin(2016)]%
        {chen2016xgboost}
\bibfield{author}{\bibinfo{person}{Tianqi Chen} {and} \bibinfo{person}{Carlos
  Guestrin}.} \bibinfo{year}{2016}\natexlab{}.
\newblock \showarticletitle{{XGBoost}: A Scalable Tree Boosting System}. In
  \bibinfo{booktitle}{\emph{Proceedings of the 22nd ACM SIGKDD International
  Conference on Knowledge Discovery and Data Mining (KDD)}}.
  \bibinfo{pages}{785--794}.
\newblock
\href{https://doi.org/10.1145/2939672.2939785}{doi:\nolinkurl{10.1145/2939672.2939785}}


\bibitem[Cho et~al\mbox{.}(2026)]%
        {cho2026librarian}
\bibfield{author}{\bibinfo{person}{Seunghyuk Cho}, \bibinfo{person}{Sunghyun
  Choi}, \bibinfo{person}{Jaeseung Heo}, \bibinfo{person}{Youngbin Choi},
  \bibinfo{person}{Saemi Moon}, \bibinfo{person}{MoonJeong Park}, {and}
  \bibinfo{person}{Dongwoo Kim}.} \bibinfo{year}{2026}\natexlab{}.
\newblock \bibinfo{title}{Long Live the Librarian! A Persistent Search
  Sub-Agent for Energy-Efficient Multi-Agent Software Engineering Systems}.
\newblock
\showeprint[arxiv]{2605.27787}~[cs.MA]
\href{https://doi.org/10.48550/arXiv.2605.27787}{doi:\nolinkurl{10.48550/arXiv.2605.27787}}
\newblock
\shownote{Accepted to EMNLP 2026 (Main Conference)}.


\bibitem[Ding et~al\mbox{.}(2024)]%
        {ding2024hybridllm}
\bibfield{author}{\bibinfo{person}{Dujian Ding}, \bibinfo{person}{Ankur
  Mallick}, \bibinfo{person}{Chi Wang}, \bibinfo{person}{Robert Sim},
  \bibinfo{person}{Subhabrata Mukherjee}, \bibinfo{person}{Victor R{\"u}hle},
  \bibinfo{person}{Laks V.~S. Lakshmanan}, {and} \bibinfo{person}{Ahmed~Hassan
  Awadallah}.} \bibinfo{year}{2024}\natexlab{}.
\newblock \showarticletitle{Hybrid {LLM}: Cost-Efficient and Quality-Aware
  Query Routing}. In \bibinfo{booktitle}{\emph{International Conference on
  Learning Representations (ICLR)}}.
\newblock


\bibitem[{Electricity Maps}(2026)]%
        {electricitymaps2026api}
\bibfield{author}{\bibinfo{person}{{Electricity Maps}}.}
  \bibinfo{year}{2026}\natexlab{}.
\newblock \bibinfo{title}{Electricity Maps {API}: Carbon Intensity Data}.
\newblock \bibinfo{howpublished}{Web API}.
\newblock
\urldef\tempurl%
\url{https://www.electricitymaps.com}
\showURL{%
\tempurl}
\newblock
\shownote{Five-minute life-cycle carbon intensity for AT, GB, BE, and
  US-CAL-CISO; data retrieved February 10, 2026}.


\bibitem[{Ember} and {Our World in Data}(2026)]%
        {ember2026carbonintensity}
\bibfield{author}{\bibinfo{person}{{Ember}} {and} \bibinfo{person}{{Our World
  in Data}}.} \bibinfo{year}{2026}\natexlab{}.
\newblock \bibinfo{title}{Lifecycle Carbon Intensity of Electricity
  Generation}.
\newblock \bibinfo{howpublished}{Our World in Data}.
\newblock
\urldef\tempurl%
\url{https://ourworldindata.org/grapher/carbon-intensity-electricity}
\showURL{%
\tempurl}
\newblock
\shownote{Ember data processed by Our World in Data; data version 2026-06-30;
  indicator 1295504; downloaded September 14, 2026}.


\bibitem[Gao et~al\mbox{.}(2024)]%
        {gao2024cachedattention}
\bibfield{author}{\bibinfo{person}{Bin Gao}, \bibinfo{person}{Zhuomin He},
  \bibinfo{person}{Puru Sharma}, \bibinfo{person}{Qingxuan Kang},
  \bibinfo{person}{Djordje Jevdjic}, \bibinfo{person}{Junbo Deng},
  \bibinfo{person}{Xingkun Yang}, \bibinfo{person}{Zhou Yu}, {and}
  \bibinfo{person}{Pengfei Zuo}.} \bibinfo{year}{2024}\natexlab{}.
\newblock \showarticletitle{Cost-Efficient Large Language Model Serving for
  Multi-turn Conversations with {CachedAttention}}. In
  \bibinfo{booktitle}{\emph{Proceedings of the 2024 USENIX Annual Technical
  Conference (USENIX ATC)}}.
\newblock


\bibitem[Golden et~al\mbox{.}(2026)]%
        {golden2026xpuathalon}
\bibfield{author}{\bibinfo{person}{Alicia Golden}, \bibinfo{person}{Carole-Jean
  Wu}, \bibinfo{person}{Gu-Yeon Wei}, {and} \bibinfo{person}{David Brooks}.}
  \bibinfo{year}{2026}\natexlab{}.
\newblock \bibinfo{title}{The {xPU}-athalon: Quantifying the Competition of
  {AI} Acceleration}.
\newblock
\showeprint[arxiv]{2604.10852}~[cs.AR]
\href{https://doi.org/10.48550/arXiv.2604.10852}{doi:\nolinkurl{10.48550/arXiv.2604.10852}}
\newblock
\shownote{Accepted to ISPASS 2026}.


\bibitem[{Google}(2026)]%
        {google2026pricing}
\bibfield{author}{\bibinfo{person}{{Google}}.} \bibinfo{year}{2026}\natexlab{}.
\newblock \bibinfo{title}{{Gemini API} Pricing}.
\newblock
  \bibinfo{howpublished}{\url{https://ai.google.dev/gemini-api/docs/pricing}}.
\newblock
\newblock
\shownote{Accessed: 2026-10-16}.


\bibitem[Hanafy et~al\mbox{.}(2023)]%
        {hanafy2024carbonscaler}
\bibfield{author}{\bibinfo{person}{Walid~A. Hanafy}, \bibinfo{person}{Qianlin
  Liang}, \bibinfo{person}{Noman Bashir}, \bibinfo{person}{David Irwin}, {and}
  \bibinfo{person}{Prashant Shenoy}.} \bibinfo{year}{2023}\natexlab{}.
\newblock \showarticletitle{{CarbonScaler}: Leveraging Cloud Workload
  Elasticity for Optimizing Carbon-Efficiency}.
\newblock \bibinfo{journal}{\emph{Proceedings of the ACM on Measurement and
  Analysis of Computing Systems}} \bibinfo{volume}{7}, \bibinfo{number}{3}
  (\bibinfo{year}{2023}), \bibinfo{pages}{1--28}.
\newblock
\href{https://doi.org/10.1145/3626788}{doi:\nolinkurl{10.1145/3626788}}


\bibitem[{International Energy Agency}(2025)]%
        {iea2025energyai}
\bibfield{author}{\bibinfo{person}{{International Energy Agency}}.}
  \bibinfo{year}{2025}\natexlab{}.
\newblock \bibinfo{booktitle}{\emph{Energy and {AI}}}.
\newblock \bibinfo{type}{{T}echnical {R}eport}.
  \bibinfo{institution}{International Energy Agency}, \bibinfo{address}{Paris}.
\newblock
\urldef\tempurl%
\url{https://www.iea.org/reports/energy-and-ai}
\showURL{%
\tempurl}


\bibitem[Jegham et~al\mbox{.}(2025)]%
        {jegham2025hungry}
\bibfield{author}{\bibinfo{person}{Nidhal Jegham}, \bibinfo{person}{Marwan
  Abdelatti}, \bibinfo{person}{Chan~Young Koh}, \bibinfo{person}{Lassad
  Elmoubarki}, {and} \bibinfo{person}{Abdeltawab Hendawi}.}
  \bibinfo{year}{2025}\natexlab{}.
\newblock \bibinfo{title}{How Hungry is {AI}? Benchmarking Energy, Water, and
  Carbon Footprint of {LLM} Inference}.
\newblock
\showeprint[arxiv]{2505.09598}~[cs.CY]
\href{https://doi.org/10.48550/arXiv.2505.09598}{doi:\nolinkurl{10.48550/arXiv.2505.09598}}


\bibitem[Kwon et~al\mbox{.}(2023)]%
        {kwon2023vllm}
\bibfield{author}{\bibinfo{person}{Woosuk Kwon}, \bibinfo{person}{Zhuohan Li},
  \bibinfo{person}{Siyuan Zhuang}, \bibinfo{person}{Ying Sheng},
  \bibinfo{person}{Lianmin Zheng}, \bibinfo{person}{Cody~Hao Yu},
  \bibinfo{person}{Joseph~E. Gonzalez}, \bibinfo{person}{Hao Zhang}, {and}
  \bibinfo{person}{Ion Stoica}.} \bibinfo{year}{2023}\natexlab{}.
\newblock \showarticletitle{Efficient Memory Management for Large Language
  Model Serving with {PagedAttention}}. In
  \bibinfo{booktitle}{\emph{Proceedings of the 29th Symposium on Operating
  Systems Principles (SOSP)}}.
\newblock


\bibitem[Li et~al\mbox{.}(2024)]%
        {li2024equitable}
\bibfield{author}{\bibinfo{person}{Pengfei Li}, \bibinfo{person}{Jianyi Yang},
  \bibinfo{person}{Adam Wierman}, {and} \bibinfo{person}{Shaolei Ren}.}
  \bibinfo{year}{2024}\natexlab{}.
\newblock \showarticletitle{Towards Environmentally Equitable {AI} via
  Geographical Load Balancing}. In \bibinfo{booktitle}{\emph{Proceedings of the
  15th ACM International Conference on Future and Sustainable Energy Systems
  (e-Energy)}}.
\newblock


\bibitem[Li et~al\mbox{.}(2026)]%
        {li2026faircarbon}
\bibfield{author}{\bibinfo{person}{Tingting Li}, \bibinfo{person}{Ziming Zhao},
  \bibinfo{person}{Zhaoxuan Li}, \bibinfo{person}{Xiaofei Yue}, {and}
  \bibinfo{person}{Jiongchi Yu}.} \bibinfo{year}{2026}\natexlab{}.
\newblock \showarticletitle{Fair and Carbon-Aware {LLM} Routing for Web
  Services}. In \bibinfo{booktitle}{\emph{Proceedings of the ACM Web Conference
  2026 (WWW '26)}}. \bibinfo{address}{Dubai, United Arab Emirates}.
\newblock


\bibitem[Liu et~al\mbox{.}(2026)]%
        {liu2026droidspeak}
\bibfield{author}{\bibinfo{person}{Yuhan Liu}, \bibinfo{person}{Yuyang Huang},
  \bibinfo{person}{Jiayi Yao}, \bibinfo{person}{Shaoting Feng},
  \bibinfo{person}{Zhuohan Gu}, \bibinfo{person}{Kuntai Du},
  \bibinfo{person}{Hanchen Li}, \bibinfo{person}{Yihua Cheng},
  \bibinfo{person}{Junchen Jiang}, \bibinfo{person}{Shan Lu}, {et~al\mbox{.}}}
  \bibinfo{year}{2026}\natexlab{}.
\newblock \showarticletitle{{DroidSpeak}: {KV} Cache Sharing Across Fine-tuned
  Model Variants}. In \bibinfo{booktitle}{\emph{23rd USENIX Symposium on
  Networked Systems Design and Implementation (NSDI 26)}}.
  \bibinfo{pages}{319--338}.
\newblock


\bibitem[Liu et~al\mbox{.}(2011)]%
        {liu2011greening}
\bibfield{author}{\bibinfo{person}{Zhenhua Liu}, \bibinfo{person}{Minghong
  Lin}, \bibinfo{person}{Adam Wierman}, \bibinfo{person}{Steven~H. Low}, {and}
  \bibinfo{person}{Lachlan L.~H. Andrew}.} \bibinfo{year}{2011}\natexlab{}.
\newblock \showarticletitle{Greening Geographical Load Balancing}. In
  \bibinfo{booktitle}{\emph{Proceedings of the ACM SIGMETRICS International
  Conference on Measurement and Modeling of Computer Systems}}.
\newblock
\href{https://doi.org/10.1145/1993744.1993767}{doi:\nolinkurl{10.1145/1993744.1993767}}


\bibitem[More et~al\mbox{.}(2026)]%
        {more2026cedar}
\bibfield{author}{\bibinfo{person}{Amit More}, \bibinfo{person}{Tarique Anwar},
  {and} \bibinfo{person}{Poonam Yadav}.} \bibinfo{year}{2026}\natexlab{}.
\newblock \showarticletitle{{CEDAR}: Carbon Efficient Dynamic Allocation and
  Routing for Agentic {LLM} Inference}. In
  \bibinfo{booktitle}{\emph{Proceedings of the 2nd International Workshop on
  Systems and Methods for Sustainable Large-Scale AI (GreenSys '26)}}.
  \bibinfo{publisher}{ACM}, \bibinfo{pages}{25--30}.
\newblock
\href{https://doi.org/10.1145/3802973.3804457}{doi:\nolinkurl{10.1145/3802973.3804457}}


\bibitem[Niu et~al\mbox{.}(2026)]%
        {niu2026tokenpowerbench}
\bibfield{author}{\bibinfo{person}{Chenxu Niu}, \bibinfo{person}{Wei Zhang},
  \bibinfo{person}{Jie Li}, \bibinfo{person}{Yongjian Zhao},
  \bibinfo{person}{Tongyang Wang}, \bibinfo{person}{Xi Wang}, {and}
  \bibinfo{person}{Yong Chen}.} \bibinfo{year}{2026}\natexlab{}.
\newblock \showarticletitle{{TokenPowerBench}: Benchmarking the Power
  Consumption of {LLM} Inference}. In \bibinfo{booktitle}{\emph{Proceedings of
  the AAAI Conference on Artificial Intelligence (AAAI-26)}},
  Vol.~\bibinfo{volume}{40}. \bibinfo{pages}{32582--32590}.
\newblock
\href{https://doi.org/10.1609/aaai.v40i38.40535}{doi:\nolinkurl{10.1609/aaai.v40i38.40535}}


\bibitem[Ong et~al\mbox{.}(2025)]%
        {ong2025routellm}
\bibfield{author}{\bibinfo{person}{Isaac Ong}, \bibinfo{person}{Amjad
  Almahairi}, \bibinfo{person}{Vincent Wu}, \bibinfo{person}{Wei-Lin Chiang},
  \bibinfo{person}{Tianhao Wu}, \bibinfo{person}{Joseph~E. Gonzalez},
  \bibinfo{person}{M.~Waleed Kadous}, {and} \bibinfo{person}{Ion Stoica}.}
  \bibinfo{year}{2025}\natexlab{}.
\newblock \showarticletitle{{RouteLLM}: Learning to Route {LLMs} from
  Preference Data}. In \bibinfo{booktitle}{\emph{International Conference on
  Learning Representations (ICLR)}}.
\newblock


\bibitem[{OpenAI}(2026)]%
        {openai2026pricing}
\bibfield{author}{\bibinfo{person}{{OpenAI}}.} \bibinfo{year}{2026}\natexlab{}.
\newblock \bibinfo{title}{{API} Pricing}.
\newblock
  \bibinfo{howpublished}{\url{https://developers.openai.com/api/docs/pricing}}.
\newblock
\newblock
\shownote{Accessed: 2026-10-16}.


\bibitem[Ouyang et~al\mbox{.}(2026)]%
        {ouyang2026unicache}
\bibfield{author}{\bibinfo{person}{Bei Ouyang}, \bibinfo{person}{Yifan Qiao},
  {and} \bibinfo{person}{Jiarong Xing}.} \bibinfo{year}{2026}\natexlab{}.
\newblock \showarticletitle{{UniCache}: Unifying Prefix Cache Eviction for
  Heterogeneous LLM Serving Workloads}.
\newblock \bibinfo{journal}{\emph{Proceedings of the ACM on Measurement and
  Analysis of Computing Systems}} \bibinfo{volume}{10}, \bibinfo{number}{2},
  Article \bibinfo{articleno}{54} (\bibinfo{year}{2026}),
  \bibinfo{numpages}{27}~pages.
\newblock
\href{https://doi.org/10.1145/3805652}{doi:\nolinkurl{10.1145/3805652}}


\bibitem[Radovanovi{\'c} et~al\mbox{.}(2023)]%
        {radovanovic2022carbonaware}
\bibfield{author}{\bibinfo{person}{Ana Radovanovi{\'c}}, \bibinfo{person}{Ross
  Koningstein}, \bibinfo{person}{Ian Schneider}, \bibinfo{person}{Bokan Chen},
  \bibinfo{person}{Alexandre Duarte}, \bibinfo{person}{Binz Roy},
  \bibinfo{person}{Diyue Xiao}, \bibinfo{person}{Maya Haridasan},
  \bibinfo{person}{Patrick Hung}, \bibinfo{person}{Nick Care},
  \bibinfo{person}{Saurav Talukdar}, \bibinfo{person}{Eric Mullen},
  \bibinfo{person}{Kendal Smith}, \bibinfo{person}{MariEllen Cottman}, {and}
  \bibinfo{person}{Walfredo Cirne}.} \bibinfo{year}{2023}\natexlab{}.
\newblock \showarticletitle{Carbon-Aware Computing for Datacenters}.
\newblock \bibinfo{journal}{\emph{IEEE Transactions on Power Systems}}
  \bibinfo{volume}{38}, \bibinfo{number}{2} (\bibinfo{year}{2023}),
  \bibinfo{pages}{1270--1280}.
\newblock
\showeprint[arxiv]{2106.11750}
\href{https://doi.org/10.1109/TPWRS.2022.3173250}{doi:\nolinkurl{10.1109/TPWRS.2022.3173250}}


\bibitem[Sheshanarayana et~al\mbox{.}(2026)]%
        {sheshanarayana2026gar}
\bibfield{author}{\bibinfo{person}{Disha Sheshanarayana},
  \bibinfo{person}{Rajat~Subhra Pal}, \bibinfo{person}{Manjira Sinha}, {and}
  \bibinfo{person}{Tirthankar Dasgupta}.} \bibinfo{year}{2026}\natexlab{}.
\newblock \bibinfo{title}{{GAR}: Carbon-Aware Routing for {LLM} Inference via
  Constrained Optimization}.
\newblock
\showeprint[arxiv]{2605.11603}
\urldef\tempurl%
\url{https://arxiv.org/abs/2605.11603}
\showURL{%
\tempurl}


\bibitem[Srivatsa et~al\mbox{.}(2025)]%
        {srivatsa2025preble}
\bibfield{author}{\bibinfo{person}{Vikranth Srivatsa}, \bibinfo{person}{Zijian
  He}, \bibinfo{person}{Reyna Abhyankar}, \bibinfo{person}{Dongming Li}, {and}
  \bibinfo{person}{Yiying Zhang}.} \bibinfo{year}{2025}\natexlab{}.
\newblock \showarticletitle{Preble: Efficient Distributed Prompt Scheduling for
  {LLM} Serving}. In \bibinfo{booktitle}{\emph{International Conference on
  Learning Representations (ICLR)}}.
\newblock


\bibitem[Sukprasert et~al\mbox{.}(2024)]%
        {sukprasert2024limitations}
\bibfield{author}{\bibinfo{person}{Thanathorn Sukprasert},
  \bibinfo{person}{Abel Souza}, \bibinfo{person}{Noman Bashir},
  \bibinfo{person}{David Irwin}, {and} \bibinfo{person}{Prashant Shenoy}.}
  \bibinfo{year}{2024}\natexlab{}.
\newblock \showarticletitle{On the Limitations of Carbon-Aware Temporal and
  Spatial Workload Shifting in the Cloud}. In
  \bibinfo{booktitle}{\emph{Proceedings of the 19th European Conference on
  Computer Systems (EuroSys)}}.
\newblock


\bibitem[Tian et~al\mbox{.}(2026)]%
        {tian2025greencache}
\bibfield{author}{\bibinfo{person}{Yuyang Tian}, \bibinfo{person}{Desen Sun},
  \bibinfo{person}{Yi Ding}, {and} \bibinfo{person}{Sihang Liu}.}
  \bibinfo{year}{2026}\natexlab{}.
\newblock \showarticletitle{Cache Your Prompt When It's Green: Carbon-Aware
  Caching for Large Language Model Serving}.
\newblock \bibinfo{journal}{\emph{Proceedings of the ACM on Measurement and
  Analysis of Computing Systems}} \bibinfo{volume}{10}, \bibinfo{number}{1}
  (\bibinfo{year}{2026}).
\newblock
\href{https://doi.org/10.1145/3788087}{doi:\nolinkurl{10.1145/3788087}}


\bibitem[Toniolo et~al\mbox{.}(2026)]%
        {toniolo2026gorgo}
\bibfield{author}{\bibinfo{person}{Alessio~Ricci Toniolo},
  \bibinfo{person}{Abinaya Dinesh}, {and} \bibinfo{person}{Rome Thorstenson}.}
  \bibinfo{year}{2026}\natexlab{}.
\newblock \bibinfo{title}{{GORGO}: Maximizing {KV}-Cache Reuse While Minimizing
  Network Latency in Cross-Region {LLM} Load Balancing}.
\newblock
\showeprint[arxiv]{2602.11688v1}
\urldef\tempurl%
\url{https://arxiv.org/abs/2602.11688v1}
\showURL{%
\tempurl}


\bibitem[{UCF-ML-Research}(2026)]%
        {r2router2026release}
\bibfield{author}{\bibinfo{person}{{UCF-ML-Research}}.}
  \bibinfo{year}{2026}\natexlab{}.
\newblock \bibinfo{title}{{R2-Router}: Released Predictors and Model
  Configuration}.
\newblock \bibinfo{howpublished}{GitHub repository}.
\newblock
\urldef\tempurl%
\url{https://github.com/UCF-ML-Research/R2-Router}
\showURL{%
\tempurl}
\newblock
\shownote{README and r2\_router/config.json; accessed September 14, 2026}.


\bibitem[Wang et~al\mbox{.}(2025)]%
        {wang2025characterization}
\bibfield{author}{\bibinfo{person}{Haonan Wang}, \bibinfo{person}{Xuxin Xiao},
  \bibinfo{person}{Mingyu Yan}, \bibinfo{person}{Zhuoyuan Zhu},
  \bibinfo{person}{Dengke Han}, \bibinfo{person}{Duo Wang},
  \bibinfo{person}{Wenming Li}, \bibinfo{person}{Xiaochun Ye},
  \bibinfo{person}{Cunchen Hu}, \bibinfo{person}{Hongyang Chen}, {and}
  \bibinfo{person}{Guangyu Sun}.} \bibinfo{year}{2025}\natexlab{}.
\newblock \bibinfo{title}{A Systematic Characterization of {LLM} Inference on
  {GPUs}}.
\newblock
\showeprint[arxiv]{2512.01644}~[cs.AR]
\href{https://doi.org/10.48550/arXiv.2512.01644}{doi:\nolinkurl{10.48550/arXiv.2512.01644}}


\bibitem[Wiesner et~al\mbox{.}(2021)]%
        {wiesner2021letswait}
\bibfield{author}{\bibinfo{person}{Philipp Wiesner}, \bibinfo{person}{Ilja
  Behnke}, \bibinfo{person}{Dominik Scheinert}, \bibinfo{person}{Kordian
  Gontarska}, {and} \bibinfo{person}{Lauritz Thamsen}.}
  \bibinfo{year}{2021}\natexlab{}.
\newblock \showarticletitle{Let's Wait Awhile: How Temporal Workload Shifting
  Can Reduce Carbon Emissions in the Cloud}. In
  \bibinfo{booktitle}{\emph{Proceedings of the 22nd International Middleware
  Conference (Middleware)}}.
\newblock
\href{https://doi.org/10.1145/3464298.3493399}{doi:\nolinkurl{10.1145/3464298.3493399}}


\bibitem[Wilkins et~al\mbox{.}(2026)]%
        {wilkins2026servers}
\bibfield{author}{\bibinfo{person}{Grant Wilkins}, \bibinfo{person}{Fiodar
  Kazhamiaka}, {and} \bibinfo{person}{Ram Rajagopal}.}
  \bibinfo{year}{2026}\natexlab{}.
\newblock \bibinfo{title}{From Servers to Sites: Compositional Power Trace
  Generation of {LLM} Inference for Infrastructure Planning}.
\newblock
\showeprint[arxiv]{2603.18383}~[cs.DC]
\href{https://doi.org/10.48550/arXiv.2603.18383}{doi:\nolinkurl{10.48550/arXiv.2603.18383}}


\bibitem[Wu et~al\mbox{.}(2026)]%
        {wu2026energylaw}
\bibfield{author}{\bibinfo{person}{Vince Wu}, \bibinfo{person}{Jiaqi Han},
  \bibinfo{person}{Boris Ruf}, \bibinfo{person}{Vincent Grari}, {and}
  \bibinfo{person}{Marcin Detyniecki}.} \bibinfo{year}{2026}\natexlab{}.
\newblock \showarticletitle{Prefill, Decode, and Context: A Measured Energy Law
  for Agentic {LLM} Serving}. In \bibinfo{booktitle}{\emph{NeurIPS 2026
  Workshop on Resource-Aware Agentic {AI}}}.
\newblock
\urldef\tempurl%
\url{https://openreview.net/forum?id=IvZ8KlwQfb}
\showURL{%
\tempurl}


\bibitem[Xia et~al\mbox{.}(2026)]%
        {xia2026skywalker}
\bibfield{author}{\bibinfo{person}{Tian Xia}, \bibinfo{person}{Ziming Mao},
  \bibinfo{person}{Jamison Kerney}, \bibinfo{person}{Ethan~J. Jackson},
  \bibinfo{person}{Zhifei Li}, \bibinfo{person}{Jiarong Xing},
  \bibinfo{person}{Scott Shenker}, {and} \bibinfo{person}{Ion Stoica}.}
  \bibinfo{year}{2026}\natexlab{}.
\newblock \showarticletitle{{SkyWalker}: A Locality-Aware Cross-Region Load
  Balancer for {LLM} Inference}. In \bibinfo{booktitle}{\emph{Proceedings of
  the 21st European Conference on Computer Systems (EuroSys)}}.
\newblock


\bibitem[Xue et~al\mbox{.}(2026)]%
        {xue2026r2router}
\bibfield{author}{\bibinfo{person}{Jiaqi Xue}, \bibinfo{person}{Qian Lou},
  \bibinfo{person}{Jiarong Xing}, {and} \bibinfo{person}{Heng Huang}.}
  \bibinfo{year}{2026}\natexlab{}.
\newblock \bibinfo{title}{{R2-Router}: A New Paradigm for {LLM} Routing with
  Reasoning}.
\newblock
\showeprint[arxiv]{2602.02823}~[cs.CL]
\href{https://doi.org/10.48550/arXiv.2602.02823}{doi:\nolinkurl{10.48550/arXiv.2602.02823}}
\newblock
\shownote{Version 2}.


\bibitem[Yan et~al\mbox{.}(2026)]%
        {yan2026agentdecarbonizer}
\bibfield{author}{\bibinfo{person}{Leyi Yan}, \bibinfo{person}{Shuangning Li},
  {and} \bibinfo{person}{Sihang Liu}.} \bibinfo{year}{2026}\natexlab{}.
\newblock \bibinfo{title}{{AgentDecarbonizer}: Carbon-Aware Execution for {AI}
  Agents}.
\newblock
\showeprint[arxiv]{2608.20566}~[cs.LG]
\href{https://doi.org/10.48550/arXiv.2608.20566}{doi:\nolinkurl{10.48550/arXiv.2608.20566}}


\bibitem[Yan et~al\mbox{.}(2025)]%
        {yan2026sharechat}
\bibfield{author}{\bibinfo{person}{Yueru Yan}, \bibinfo{person}{Tuc Nguyen},
  \bibinfo{person}{Bo Su}, \bibinfo{person}{Melissa Lieffers}, {and}
  \bibinfo{person}{Thai Le}.} \bibinfo{year}{2025}\natexlab{}.
\newblock \bibinfo{title}{{ShareChat}: A Dataset of Chatbot Conversations in
  the Wild}.
\newblock
\showeprint[arxiv]{2512.17843}~[cs.CL]
\href{https://doi.org/10.48550/arXiv.2512.17843}{doi:\nolinkurl{10.48550/arXiv.2512.17843}}


\bibitem[Zheng et~al\mbox{.}(2024)]%
        {zheng2024sglang}
\bibfield{author}{\bibinfo{person}{Lianmin Zheng}, \bibinfo{person}{Liangsheng
  Yin}, \bibinfo{person}{Zhiqiang Xie}, \bibinfo{person}{Chuyue Sun},
  \bibinfo{person}{Jeff Huang}, \bibinfo{person}{Cody~Hao Yu},
  \bibinfo{person}{Shiyi Cao}, \bibinfo{person}{Christos Kozyrakis},
  \bibinfo{person}{Ion Stoica}, \bibinfo{person}{Joseph~E. Gonzalez},
  \bibinfo{person}{Clark Barrett}, {and} \bibinfo{person}{Ying Sheng}.}
  \bibinfo{year}{2024}\natexlab{}.
\newblock \showarticletitle{{SGLang}: Efficient Execution of Structured
  Language Model Programs}. In \bibinfo{booktitle}{\emph{Advances in Neural
  Information Processing Systems (NeurIPS)}}.
\newblock


\end{thebibliography}

\clearpage
\appendix

\section{Energy and Latency Profiles}
\label{sec:impl-params}
\label{sec:latency-profiles}

\noindent\textbf{Energy and prefill.} An uncached input token of a model with $P_m$ parameters costs $2P_m$ FLOPs. With prefill utilization (MFU) $\eta_{m,h}$, per-GPU peak FP16 rate $F_h$, prefill power $f\,\mathrm{TDP}_h$ per GPU, and $G_{m,h}$ GPUs, its energy and the prefill throughput are
\begin{equation}
\begin{aligned}
e^{\mathrm{in}}_{m,h}&=\frac{2P_m\,f\,\mathrm{TDP}_h}{\eta_{m,h}F_h},
\qquad
v^{\mathrm{pre}}_{m,h}=\frac{\eta_{m,h}G_{m,h}F_h}{2P_m},\\
\eta_{m,h}&=\eta_{\mathrm{ref}}\left(\frac{P_m}{70\mathrm{B}}\right)^{1-\alpha}\eta_h.
\end{aligned}
\label{eq:energy-model}
\end{equation}
$F_h$ is 125, 312, and 989.5~TFLOP/s and $\mathrm{TDP}_h$ is 150, 400, and 700~W for A10, A100-40GB, and H100-SXM. We set $f=0.85$, within the measured 0.75--1.0 of TDP~\cite{golden2026xpuathalon,wilkins2026servers}, and $\eta_{\mathrm{ref}}=0.5$ for Llama-3.1-70B on eight A100s. The exponent $\alpha=0.8$ lies between the size scaling of energy per token measured on H100 (about 0.5)~\cite{niu2026tokenpowerbench} and on A100 (about 0.9)~\cite{wang2025characterization}. The GPU factors $\eta_h$ (0.653, 1, and 0.69) keep the relative energies of 1.433, 1, and 0.8 on A10, A100, and H100 of the 70B reference profile of Li et al.~\cite{li2026faircarbon}, and the H100 value is close to the measured H100-to-A100 energy per FLOP of 0.74~\cite{antepara2025energy}. Output and cached input tokens cost $e^{\mathrm{out}}=\kappa\,e^{\mathrm{in}}$ and $e^{\mathrm{cache}}=\gamma\,e^{\mathrm{in}}$.

\noindent\textbf{Reference energy.} For Llama-3.1-70B on eight A100s, Equation~\eqref{eq:energy-model} gives $e^{\mathrm{in}}_{\mathrm{ref}}=8.48\times10^{-8}$~kWh (0.305~J) per token, so $e^{\mathrm{out}}_{\mathrm{ref}}=5.09\times10^{-7}$ and $e^{\mathrm{cache}}_{\mathrm{ref}}=4.24\times10^{-9}$~kWh per token. Table~\ref{tab:profiles} lists the energy of every deployment relative to this reference, $\rho_{m,h}=e^{\mathrm{in}}_{m,h}/e^{\mathrm{in}}_{\mathrm{ref}}$. Because $f$ and $\eta_{\mathrm{ref}}$ cancel in $\rho_{m,h}$, they do not affect which deployment has the lowest estimated carbon.

\noindent\textbf{Decoding.} A batch-one decode step is memory-bound and reads the 16-bit weights once. With per-GPU bandwidth $B_h$ of 600, 1,555, and 3,350~GB/s, the decoding throughput is $v^{\mathrm{dec}}_{m,h}=G_{m,h}B_h/(2P_m)$. With $\kappa=6$, the energy of an output token corresponds to an effective decode batch of about 5--17 sequences. A memory-bound step changes little up to this batch size, so we use the batch-one decoding time.

\noindent\textbf{Deployments.} Each model runs on one GPU except Llama-3.1-70B, which uses eight A10, eight A100-40GB, or four H100 GPUs. These counts divide its 64 query heads and 8 KV heads, as tensor parallelism requires. With 90\% of GPU memory usable, they leave 32--147~GB for the KV cache after its 141~GB of 16-bit weights, whereas four A100-40GB or two H100 GPUs would leave about 3~GB.

\begin{table}[t]
\centering
\caption{Analytical model--GPU profiles. Energy is relative to Llama-3.1-70B on A100, and throughputs are in tokens/s.}
\label{tab:profiles}
\small
\setlength{\tabcolsep}{4pt}
\begin{tabular}{llrrrr}
\toprule
Model & GPU & \twoline{Relative\\energy} & Prefill & MFU & Decode\\
\midrule
\multirow{3}{*}{Gemma-3-1B} & A10 & 0.048 & 8,725 & 0.14 & 300.0\\
 & A100 & 0.033 & 33,348 & 0.21 & 777.5\\
 & H100 & 0.027 & 72,950 & 0.15 & 1,675.0\\
\midrule
\multirow{3}{*}{Llama-3.2-3B} & A10 & 0.115 & 3,623 & 0.17 & 100.0\\
 & A100 & 0.080 & 13,848 & 0.27 & 259.2\\
 & H100 & 0.064 & 30,292 & 0.18 & 558.3\\
\midrule
\multirow{3}{*}{Gemma-3-4B} & A10 & 0.145 & 2,878 & 0.18 & 75.0\\
 & A100 & 0.101 & 11,001 & 0.28 & 194.4\\
 & H100 & 0.081 & 24,064 & 0.19 & 418.8\\
\midrule
\multirow{3}{*}{Mistral-7B-v0.2} & A10 & 0.227 & 1,839 & 0.21 & 42.9\\
 & A100 & 0.158 & 7,031 & 0.32 & 111.1\\
 & H100 & 0.127 & 15,380 & 0.22 & 239.3\\
\midrule
\multirow{3}{*}{Llama-3.1-70B} & $8\times$A10 & 1.433 & 2,332 & 0.33 & 34.3\\
 & $8\times$A100 & 1.000 & 8,914 & 0.50 & 88.9\\
 & $4\times$H100 & 0.800 & 9,750 & 0.34 & 95.7\\
\bottomrule
\end{tabular}
\end{table}

\end{document}